\documentclass[aps, prd, twocolumn, floatfix, eqsecnum, superscriptaddress,longbibliography]{revtex4-1}

\usepackage{amsmath}
\usepackage{amsfonts,amssymb, graphicx}
\usepackage{wrapfig}
\usepackage{afterpage}
\usepackage{natbib}
\usepackage{color}
\usepackage[usenames,dvipsnames,svgnames,table]{xcolor}
\usepackage{amsmath}
\usepackage{amssymb}
\usepackage{graphicx}
\usepackage[usenames,dvipsnames]{xcolor}
\usepackage{tikz}
\usepackage{pgffor}
\usepackage{verbatim}
\usepackage{float}
\usepackage{mathrsfs}
\usepackage{comment}
\usepackage{enumerate}
\usepackage[colorlinks, breaklinks=true,linkcolor=red, citecolor=blue, linktocpage=true]{hyperref}

\newcommand{\sgn}{\text{sgn}}

\renewcommand{\>}{\rangle}
\newcommand{\be}{\begin{equation} }
\newcommand{\ee}{\end{equation} }
\newcommand{\ba}{\begin{eqnarray} }
\newcommand{\ea}{\end{eqnarray} }

\newcommand{\bpm}{\begin{pmatrix}}
\newcommand{\epm}{\end{pmatrix}}
\newcommand{\bmm}{\begin{matrix}}
\newcommand{\emm}{\end{matrix}}

\newcommand{\nn}{\nonumber}

\newcommand{\la}{\label}
\newcommand{\bea}{\begin{eqnarray}}
\newcommand{\eea}{\end{eqnarray}}

\newtheorem{lemma}{Lemma}

\begin{document}

\title{Ungappable edge theories with finite dimensional Hilbert spaces}
\author{Sriram Ganeshan}
 \affiliation{Physics Department, City College of New York, NY 10031}
 \affiliation{CUNY Graduate Center, New York NY, 10031, USA}
 \author{Michael Levin}
  \affiliation{Department of Physics, Kadanoff Center for Theoretical Physics, University of Chicago, Chicago, Illinois 60637, USA}
\begin{abstract}
We construct a new class of edge theories for a family of fermionic Abelian topological phases with $K$-matrices of the form $K = \bpm k_1 & 0 \\ 0 & - k_2 \epm$, where $k_1, k_2 > 0$ are odd integers. Our edge theories are notable for two reasons: (i) they have finite dimensional Hilbert spaces (for finite sized systems) and (ii) depending on the values of $k_1, k_2$, some of the edge theories describe boundaries that cannot be gapped by any local interaction. The simplest example of such an ungappable boundary occurs for $(k_1, k_2) = (1, 3)$, which is realized by the $\nu = 2/3$ FQH state. We derive our edge theories by starting with the standard chiral boson edge theory, consisting of two counterpropagating chiral boson modes, and then introducing an array of pointlike impurity scatterers. We solve this impurity model exactly in the limit of infinite impurity scattering, and we show that the energy spectrum consists of a gapped phonon spectrum together with a ground state degeneracy that scales exponentially with the number of impurities. This ground state subspace forms the Hilbert space for our edge theory. We believe that similar edge theories can be constructed for any Abelian topological phase with vanishing thermal Hall coefficient, $\kappa_H = 0$.
 \end{abstract}
\maketitle

\section{Introduction}

Many insights into two dimensional topological phases can be obtained by examining the properties of their spatial boundaries~\cite{wen1995topological}. These boundaries are particularly interesting in cases where they host \emph{gapless} edge excitations. The structure of these edge modes can then be conveniently described in terms of an edge theory.

At its core, an ``edge theory'' for a 2D topological phase consists of two pieces of data: (i) a Hilbert space $\mathcal{H}$ and (ii) a complete list of local operators $\{\mathcal{O}\}$ acting in $\mathcal{H}$. These two pieces of data have a simple physical meaning: $\mathcal{H}$ describes the subspace of low energy edge excitations of some 2D system, while $\{\mathcal{O}\}$ describes the projections of local operators in the 2D system into this low energy subspace. We note that edge theories of systems with global symmetries carry additional structure\footnote{This structure consists of a representation of the symmetry group $G$ acting on $\mathcal{H}$.}, but in this paper we will not be interested in such ``symmetry-enriched'' topological phases.

\begin{figure}[t]
  \centering
\includegraphics[width=0.9\columnwidth]{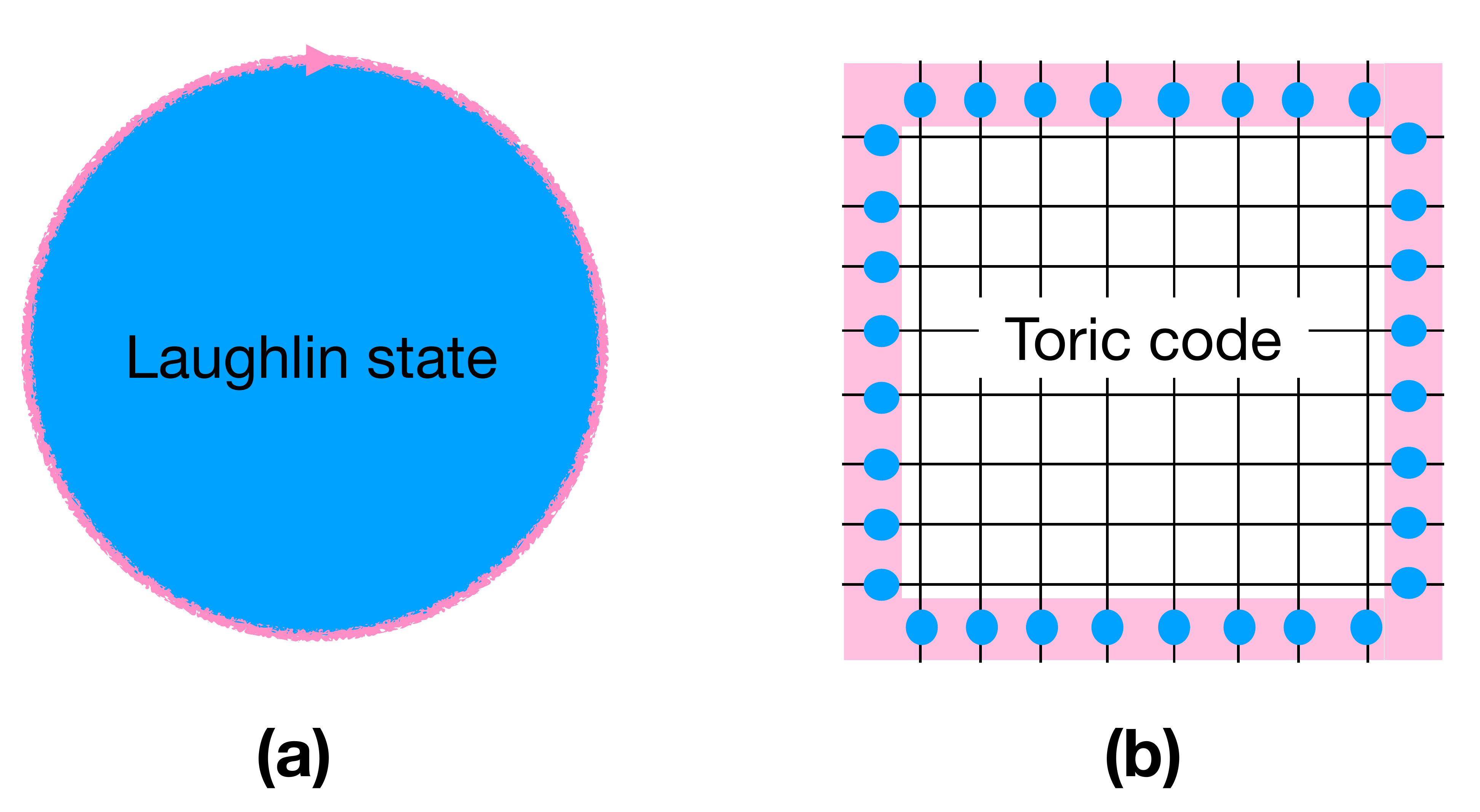}
    \caption{Examples of edge theories for type-I and type-III topological phases. (a) A chiral boson field theory describes the edge of the (type-III) Laughlin state. (b) A spin-$1/2$ chain with a global constraint describes the edge of the (type-I) toric code model. The edge degrees of freedom correspond to the spins (blue dots) along the boundary of the lattice.}
\label{fig:edgetypes}
\end{figure}

A famous example of an edge theory is the chiral boson field theory that describes the edge of $\nu = 1/k$ Laughlin fractional quantum Hall (FQH) state~\cite{wen1995topological} (Fig.~\ref{fig:edgetypes}a). This edge theory consists of a single 1D field $\phi(x)$ obeying the commutation algebra $[\phi(x), \partial_y \phi(y)] = -\frac{2\pi i}{k} \delta(x-y)$ as well as the global constraint $e^{i \int_0^L \partial_x\phi\, dx}=1$. The Hilbert space $\mathcal{H}$ is the unique irreducible representation of this operator algebra, while the local operators $\{\mathcal{O}\}$ consist of arbitrary derivatives and products of the electron creation/annihilation operators $e^{\pm i k \phi}$.

The goal of this paper is to construct edge theories that are fundamentally different from the above Laughlin edge theory. In particular we wish to find edge theories that have a \emph{finite} dimensional Hilbert space $\mathcal{H}$ for a finite size system. Such edge theories are desirable because they provide a simple and well-regulated setting to study edge physics. 

The simplest example of a finite dimensional edge theory is the ``lattice edge theory'' for the toric code model~\cite{KitaevToric, yang2014edgepeps, levin2018talkorder, ji2019noninvertible} (Fig.~\ref{fig:edgetypes}b). The Hilbert space $\mathcal{H}$ for this edge theory consists of a chain of $N$ spin-$1/2$'s, arranged in a ring, with a $\mathbb{Z}_2$ global constraint $\prod_{i=1}^N \sigma^x_i=1$. The local operators $\{\mathcal{O}\}$ consist of arbitrary products of the spin operators $\{\sigma^x_i, \sigma^z_i \sigma^z_{i+1}\}$ acting on nearby lattice sites. This edge theory describes a particular boundary of the toric code model that has $2^{N-1}$ zero energy edge states below a bulk gap.

More generally, it is natural to ask which other topological phases can support finite dimensional edge theories. To answer this question, it is useful to divide topological phases into three classes:
\begin{enumerate}[(I)]
\item{Topological phases that have a vanishing thermal Hall coefficient~\cite{kane1997thermal}, $\kappa_H = 0$, and support a gapped boundary.}
\item{Topological phases that have a vanishing thermal Hall coefficient, $\kappa_H = 0$, but do not support a gapped boundary.}
\item{Topological phases with a nonzero thermal Hall coefficient,  $\kappa_H \neq 0$.} 
\end{enumerate}

Type-I topological phases almost certainly have finite dimensional edge theories. Indeed, it is believed~\cite{kitaev2012models, lin2014stringnet, kong2014condensation, freed2020gapped} that all phases of this type can be realized by string-net models~\cite{LevinWenstrnet}\footnote{Strictly speaking, this statement is for \emph{bosonic} topological phases, but we expect that a similar statement holds in the fermionic case.}, and every string-net model has a finite dimensional edge theory similar to that of the toric code model~\cite{levinunpublished}.
On the opposite end of the spectrum are type-III topological phases: for phases in this class, it seems unlikely that a finite dimensional edge theory is possible at all, since the nonzero thermal Hall coefficient $\kappa_H \neq 0$ presumably means that the edge spectrum must form a continuum. 

The key question is then whether there exist finite dimensional edge theories for \emph{type-II} topological phases. In this paper, we answer this question in the affirmative. Our main result is a collection of finite dimensional edge theories for a family of type-I and type-II Abelian topological phases. Specifically, we construct edge theories for fermionic Abelian topological phases described by $2 \times 2$ $K$-matrices of the form $K = \bpm k_1 & 0 \\ 0 & -k_2 \epm$ where $k_1, k_2 > 0$ are odd integers. Two prototypical examples are $(k_1, k_2) = (1, 3)$ and $(k_1, k_2) =  (1, 9)$. The former is a type-II topological phase realized by the $\nu = 2/3$ FQH state, while the latter is a type-I topological phase realized by the $\nu = 8/9$ FQH state. More generally, the phases for which $k_1 \cdot k_2$ is a perfect square are of type-I, while the other phases are of type-II~\cite{kapustin2011topological, levin2013protected}.

To derive our finite dimensional edge theories, we start with the standard edge theory for $K = \bpm k_1 & 0 \\ 0 &  -k_2 \epm$, which consists of two counterpropagating chiral boson edge modes. We then introduce an array of pointlike ``impurity'' scatterers that scatter electrons between the two edge modes. The impurities we introduce come in two types: ``conventional'' impurities that scatter a single electron from one mode to the other, and ``superconducting'' impurities that scatter a single electron from one mode to a hole on the other, with a Cooper pair entering the superconductor. Using the formalism developed in an earlier work~\cite{ganeshan2016formalism}, we solve this impurity model exactly in the limit of infinitely strong impurity scattering. In this limit, the energy spectrum of the impurity model can be cleanly separated into two pieces (Fig.~\ref{fig:spsch}): (i) a gapped phonon spectrum and (ii) a ground state degeneracy that scales exponentially with the number of impurities. This ground state space forms the Hilbert space for our low energy edge theory. 

It is natural to compare our impurity model with the models considered in Refs.~\onlinecite{lindner2012fractionalizing, clarke2013exotic, cheng2012superconducting}, in which a fractional quantum spin Hall edge is proximity coupled to an alternating sequence of superconductors and ferromagnets. At first glance, the models share similar physics: like our impurity model, the models of  Refs.~\onlinecite{lindner2012fractionalizing, clarke2013exotic, cheng2012superconducting} exhibit a gapped phonon spectrum and an exponentially large ground state degeneracy. However, this analogy is (mostly) misleading. The key distinction is that the models considered in Refs.~\onlinecite{lindner2012fractionalizing, clarke2013exotic, cheng2012superconducting} describe an alternating sequence of two different types of gapped boundaries, while there is no such picture for the impurity model for general $k_1, k_2$. Relatedly, while the ground state degeneracy in Refs.~\onlinecite{lindner2012fractionalizing, clarke2013exotic, cheng2012superconducting} is topologically protected, this is \emph{not} the case for the impurity model, where the degeneracy splits at any finite $U$. The one exception to these statements is the special case $k_1 = k_2$: in that case, the impurity model is indeed a close cousin of the models considered in Refs.~\onlinecite{lindner2012fractionalizing, clarke2013exotic, cheng2012superconducting} and the analogy is valid. (See Sec.~\ref{sec:model} for more details). 

In addition to constructing edge theories, we also investigate the ``gappability'' of our edge theories in several examples. As one might expect, we are able to find gapping Hamiltonians for the type-I examples but not for the type-II examples. First, we study one of the simplest type-I edge theories, namely $(k_1, k_2) = (1,9)$. In this case, we construct a gapping Hamiltonian which is a sum of commuting local terms. Next, we consider one of the simplest type-II edge theories, namely $(k_1, k_2) = (1,3)$. In this case, we show that there is an obstruction to finding a commuting gapping Hamiltonian. We also numerically study a simple non-commuting edge Hamiltonian in this case and show that it has a gapless spectrum.

This paper is organized as follows. In Sec.~\ref{sec:model}, we present the impurity model that underlies our edge theory. In Sec.~\ref{sec:stmodel}, we solve the impurity model, and in Sec.~\ref{sec:edgeth} we derive our edge theory. In Sec.~\ref{sec:gap89}, we construct a  gapping Hamiltonian for one of the simplest type-I edge theories, namely $(k_1, k_2) = (1,9)$. In Sec.~\ref{sec:obstg}, we investigate the obstructions to gapping one of the simplest type-II edge theories, namely $(k_1, k_2) = (1,3)$. In Sec.~\ref{sec:disc}, we summarize our results and discuss extensions and future directions. Technical details are presented in the Appendix.

\section{Impurity model}
\label{sec:model}

Our goal is to derive finite dimensional edge theories for fermionic Abelian topological phases described by the $2 \times 2$ K-matrix $K = \bpm k_1 & 0 \\ 0 & - k_2 \epm$ where $k_1, k_2 > 0$ are odd integers. To do this, we start with the standard edge theory for these phases and then modify this edge theory by introducing a set of impurity scatterers. The ground state subspace of this impurity model will define our finite dimensional edge theory. In this section we describe the impurity model.

First we recall the standard edge theory for the Abelian topological phase with K-matrix $K = \bpm k_1 & 0 \\ 0 &  -k_2 \epm$. This edge theory
consists of two counterpropagating chiral boson edge modes, described by bosonic fields $\phi_{1}, \phi_2$ satisfying the following commutation relations~\cite{wen1995topological} :
\begin{align}
[\phi_1(x), \partial_y \phi_1(y)]&=-\frac{2\pi i}{k_1}\delta(x-y)\nonumber \\
[\phi_2(x),\partial_y \phi_2(y)]&=\frac{2\pi i}{k_2}\delta(x-y)\nonumber \\
[\phi_1(x),\partial_y \phi_2(y)]&=0.
 \label{eq:comm}
\end{align}
Here we use a normalization convention where the electron operators $\psi_1^\dagger, \psi_2^\dagger$ for the two edge modes are
\begin{align}
\psi_1^{\dagger}=e^{-ik_1\phi_1}, \ \ \ \psi_2^{\dagger}=e^{ik_2\phi_2}
\label{elecdef}
\end{align}
The Hamiltonian is
\begin{equation}\label{eq:hzero}
H_0 = \frac{v}{4\pi}\int_{0}^{L}[k_1(\partial_{x}\phi_{1}(x))^{2}+k_2(\partial_{x}\phi_{1}(x))^{2}]dx.
\end{equation}
where $L$ is the circumference of the (circular) edge and $v$ is the velocity of the edge modes (we choose $v$ to be the same for both modes for simplicity). 

The above theory describes a \emph{clean} edge with no scattering between the two modes. As such, there are two separately conserved $U(1)$ charges $Q_1, Q_2$ associated with the two edge modes:
\begin{align}
Q_i &= \frac{1}{2\pi} \int_{0}^{L} \partial_x \phi_i dx, \quad i = 1,2
\label{Qidef}
\end{align}
We now break both of these $U(1)$ symmetries by introducing two types of scattering terms into the Hamiltonian: a ``conventional'' scattering term that backscatters electrons from one edge mode to the other and a ``superconducting'' term that scatters an electron from one edge mode to a hole on the other mode. Given our definition of the electron operators in (\ref{elecdef}), these two types of scattering terms take the form 
\begin{align*}
U \cos(k_1 \phi_1 \pm k_2 \phi_2-\alpha)
\end{align*}
 where $U, \alpha$ describe the magnitude and phase of the impurity scattering.

\begin{figure}[t]
  \centering
\includegraphics[width=0.55\columnwidth]{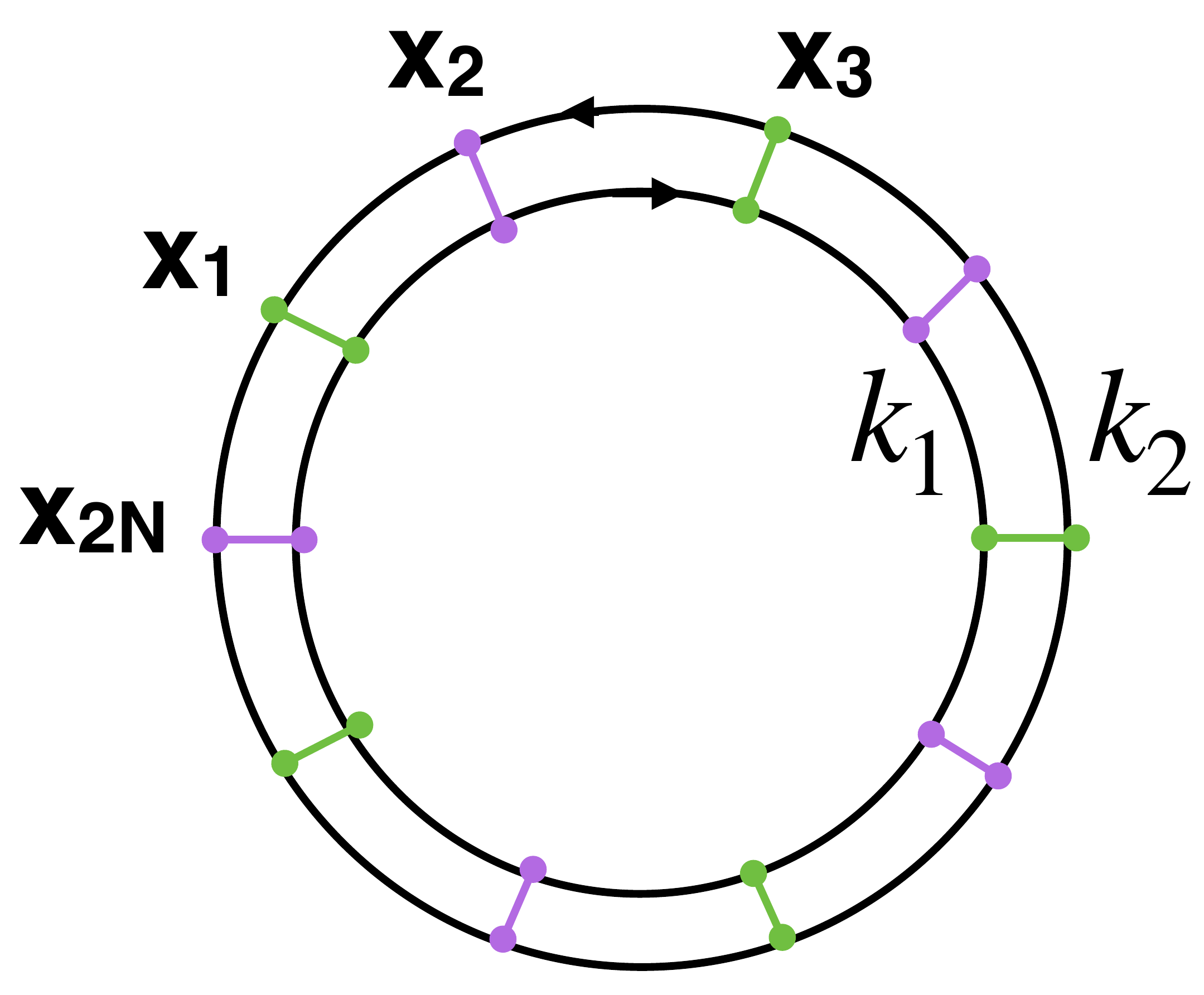}
    \caption{Impurity model: Two counterpropagating chiral boson modes with parameters $k_1, k_2$ with an alternating sequence of $2N$ impurity scatterers. Conventional impurities are located at positions $x_1,x_3,....,x_{2N-1}$; superconducting impurities  are located at positions $x_2,x_4,...,x_{2N}$.}
\la{fig:fmscmi}
\end{figure}

In order to facilitate the solution of our model, we will only introduce scattering at a set of \emph{discrete} points, $\{x_i\}$ along the edge. We think of these pointlike scatterers as describing two types of impurities: conventional impurities and superconducting impurities. We arrange the impurities in an alternating pattern with conventional impurities at positions $x_1,x_3,...,x_{2N-1}$ and superconducting impurities located at positions $x_2,x_4,...,x_{2N}$  (Fig. \ref{fig:fmscmi}). We choose the $x_l$ to be regularly spaced with a spacing $s = L/2N$, and we set all the phases $\alpha = 0$ for simplicity, which gives the following impurity model:
\begin{align}
H &=H_0-U\sum^{2N}_{l=1}\cos[k_{1}\phi_{1}(l s)+(-1)^{l+1}k_{2}\phi_{2}(l s)]\la{Hfmsc}
\end{align}

A few comments about this model: first, we should explain why we use an impurity model with \emph{two} types of impurities, instead of just one. The reason we construct our model in this way is that one type of impurity is not sufficient to open a gap in the phonon spectrum for general $k_1, k_2$. For example, suppose we only included conventional impurities, i.e.~terms of the form $U \cos(k_1 \phi_1 + k_2 \phi_2-\alpha)$. Consider the case $(k_1, k_2) = (1,3)$, which describes the edge of the $\nu = 2/3$ FQH state. In this case, it is known that a model with only conventional impurity scattering hosts two gapless counterpropagating edge modes~\cite{kane1994randomness, heinrich2017solvable}. Indeed, a gapless phonon spectrum is guaranteed, on general grounds, due to two properties of this model: (1) the total charge $Q = Q_1 + Q_2$ is conserved, and (2) the bulk topological phase has a nonzero electric Hall conductance, $\sigma_H = 2/3$. (Any system with a conserved charge and a nonzero Hall conductance must have gapless phonon excitations). Generalizing this argument, one can show\footnote{See App.~B of Ref.~\onlinecite{levin2013protected}.} that the only case in which a gap can be opened up by a single type of scattering term of the form $\cos(\Lambda_1 k_1 \phi_1 + \Lambda_2 k_2 \phi_2)$, with $\Lambda_1, \Lambda_2 \in \mathbb{Z}$, is if $\Lambda_1, \Lambda_2$ obey the ``null vector criterion'':~\cite{Haldane1995null} 
\begin{align}
k_1 \Lambda_1^2 - k_2 \Lambda_2^2 = 0.
\label{nullvector}
\end{align}
If $k_1 k_2$ is not a perfect square, then this null vector equation has no integer solutions $\Lambda_1, \Lambda_2$, and therefore a single type of impurity can never open a gap in the phonon spectrum.

Our second comment is about the special case $k_1 = k_2$. This case is special because our impurity scattering terms obey the null vector condition (\ref{nullvector}) (since $\Lambda_1, \Lambda_2 = \pm 1$). Therefore when we take the limit $U \rightarrow \infty$, each impurity effectively gaps out a short segment of boundary. We can then think of the edge, as a whole, as an alternating sequence of two different types of gapped boundaries. The physics of the impurity model is then similar to the models discussed in Refs.~\onlinecite{lindner2012fractionalizing, clarke2013exotic, cheng2012superconducting} in which a fractional quantum spin Hall edge is gapped out in two different ways by proximity coupling to an alternating sequence of superconductors and ferromagnets. This connection is discussed in more detail in Ref.~\onlinecite{ganeshan2016formalism}.

Note that the $k_1 \neq k_2$ case is qualitatively different. In this case, the scattering terms do not obey (\ref{nullvector}), and therefore we cannot think of the individual impurities as gapping out short segments of boundary. The impurity model still has a phonon gap, as we will show below, but this gap has a different character because it is a \emph{collective} property of impurities, rather than the individual impurities.

\section{Solving the impurity model}
\label{sec:stmodel}
\subsection{Review of general formalism}\label{sec:rof}
Our solution of the impurity model (\ref{Hfmsc}) is based on a general formalism for solving quadratic Hamiltonians with large cosine terms, introduced in Ref.~\onlinecite{ganeshan2016formalism}. Below we briefly review some of the most important results of this formalism before turning to our specific problem.

Consider a general Hamiltonian of the form
\begin{align}
H=H_0-U\sum_{i} \cos(C_i)
\label{genh}
\end{align}
defined on some phase space $\{x_1, p_1, x_2, p_2,...\}$.  $H_0$ is a  quadratic function of position and momentum variables $\{x_1, p_1, x_2, p_2,...\}$ and the $C_i$ are linear functions of these variables. The $C_i$'s can be arbitrary except for two restrictions: (1) $\{C_1,C_2,...\}$ are linearly independent, and (2) $[C_i,C_j]$ is an integer multiple of $2\pi i$ for all $i,j$ (so that the cosine terms commute with one another). Ref.~\onlinecite{ganeshan2016formalism} showed how to find the low energy spectrum of Hamiltonians of this kind in the limit $U \rightarrow \infty$. 

The basic idea behind the analysis of Ref.~\onlinecite{ganeshan2016formalism} is that the cosine terms act as \emph{constraints} in the limit $U \rightarrow \infty$. These constraints force the arguments of the cosine terms to be locked to integer multiples of $2\pi$ at low energies. When this happens, the low energy spectrum of $H$ can be described by an effective Hamiltonian $H_{\text{eff}}$ acting within an effective Hilbert space $\mathcal{H}_{\text{eff}}$. Importantly, the effective Hamiltonian $H_{\text{eff}}$ is \emph{quadratic} and therefore can be diagonalized using elementary methods.

How do we construct the effective Hamiltonian and Hilbert space? The Hilbert space is easy: $\mathcal{H}_{\text{eff}}$ is the subspace of the original Hilbert space consisting of all states $|\psi\>$ satisfying
\begin{align}
\cos(C_i)|\psi\>=|\psi\>, \ \ \ i=1,2,...
\label{hilbeff}
\end{align}
As for the Hamiltonian, Ref.~\onlinecite{ganeshan2016formalism} described a simple recipe for simultaneously constructing and diagonalizing $H_{\text{eff}}$. The first step is to find all operators $a$ that are linear combinations of the phase space variables $x_1, p_1,...$ and that satisfy the equations
\begin{align}
[a,H_0]&=Ea+\sum_{l}\lambda_l[C_l,H_0] \la{eq:aphonon1} \\
[a,C_l]&=0, \ \ \text{for all $l$} \la{eq:cons1}
\end{align}
where $\lambda_l$ and $E$ are arbitrary scalars with $E \neq 0$. The above operators $a$ have a simple physical meaning: they describe creation or annihilation operators for the effective Hamiltonian $H_{\text{eff}}$. The scalar $E$ is the energy of the corresponding mode while the scalars $\lambda_l$ can be thought of as Lagrange multipliers associated with the constraints imposed by the cosine terms.

Once the solutions to (\ref{eq:aphonon1}-\ref{eq:cons1}) have been identified, the next step is to separate them into two classes: `annihilation operators' with $E>0$ and `creation operators' with $E<0$. If $a_1, a_2,...$ form a complete set of linearly independent annihilation operators, and $a_1^{\dagger},a_2^\dagger,...$ are the corresponding creation operators, then they should be normalized so that
\begin{align}
[a_k,a_{k'}^{\dagger}]=\delta_{kk'}, \ \ \ [a_k,a_{k'}]=[a^{\dagger}_k,a^{\dagger}_{k'}]=0
\end{align}
After these steps have been completed, the effective Hamiltonian $H_{\text{eff}}$ can be written down easily: according to Ref.~\onlinecite{ganeshan2016formalism}, $H_{\text{eff}}$ is simply given by\footnote{More precisely, Eq.~(\ref{heffdiag}) is only guaranteed to hold if we make the additional assumption that the matrix $\mathcal{Z}_{ij} = \frac{1}{2\pi i} [C_i, C_j]$ has a non-vanishing determinant. This property holds for all the systems discussed in this paper.}
\begin{equation}
H_{\text{eff}}=\sum_{k} E_k a^{\dagger}_k a_k. 
\label{heffdiag}
\end{equation}

At this point, it is tempting to conclude that the energy spectrum of $H_{\text{eff}}$ is identical to that of a collection of harmonic oscillators with frequencies $E_k$. However, this is not correct in general. Indeed, Ref.~\onlinecite{ganeshan2016formalism} showed that each occupation number eigenstate is $D$-fold degenerate where
\begin{align}
D = \sqrt{|\det(\mathcal{Z})|}
\label{degformula}
\end{align}
and where $\mathcal{Z}_{ij}$ is the commutator matrix:
\begin{align}
 \mathcal{Z}_{ij} = \frac{1}{2\pi i} [C_i, C_j].
\end{align}
An important special case of this result is that the \emph{ground state} of $H$ is $D$-fold degenerate. This ground state degeneracy will play a central role in this paper.

The intuition behind Eq.~(\ref{degformula}) is that the degeneracy arises because the arguments of the cosine terms, $C_i$, do not commute with one another; hence
to compute the degeneracy, we need to carefully analyze the commutation relations between the $C_i$'s. See Ref.~\onlinecite{ganeshan2016formalism} for more details.

\subsection{Applying the formalism to the impurity model}

We now apply the above formalism to diagonalize the Hamiltonian $H$ (\ref{Hfmsc}) in the limit $U\rightarrow \infty$. The first step is to write $H$ in the
standard form
\begin{equation}
H = H_0 - U\sum_{l=1}^{2N}\cos(C_l)
\label{eq:edgemodel}
\end{equation}
where
\begin{align}
C_l=(k_{1}\phi_{1}(l s)+(-1)^{l+1}k_{2}\phi_{2}(l s))\quad l=1,...,2N
\label{cldef}
\end{align}

According to the general formalism, the first step in analyzing the $U \rightarrow \infty$ limit is to search for all operators $a$ with the following properties. First, $a$ should be a linear combination of the phase space variables $\partial_y \phi_1$ 
and $\partial_y \phi_2$: 
\begin{align}
a=\int_{0}^{L}dy\ [f_{1}(y)\partial_{y}\phi_{1}(y)+f_{2}(y)\partial_{y}\phi_{2}(y)].
\label{genaexp}
\end{align}
Second, $a$ should obey
\begin{align}
[a,H_0]&=Ea+\sum^{2N}_{l=1}\lambda_l[C_l,H_0]\la{eq:aphonon}\\
[a, C_l]&=0, \quad l=1,...,2N \la{eq:cons}.
\end{align}
for some scalars $\lambda_l$ and $E\ne 0$. Finally, since our model has discrete translational symmetry with a unit cell of length $2s$, the functions $f_{1,2}$ should obey the Bloch condition
\begin{align}
f_1(x+2s)=e^{-i\theta}f_1(x),	\quad f_2(x+2s)=e^{-i\theta}f_2(x)\la{eq:bloch}
\end{align}
where the crystal momentum $\theta$ is defined in $[-\pi, \pi]$ and is quantized in multiples of $2\pi/N$. 

Our task is to solve Eqs.~(\ref{eq:aphonon}), (\ref{eq:cons}) and (\ref{eq:bloch}). For clarity, we present our results first and then explain the derivation. What we find is that are an infinite number of solutions to (\ref{eq:aphonon}-\ref{eq:bloch}) for each $\theta$ in $[-\pi, \pi]$. We label these solutions as $a_{m,\theta}$ and $E_{m,\theta}$ where $m$ can be thought of as a kind of band index, which runs over the set $0, \pm 1, \pm 2,...$. We find that the $a_{m,\theta}$ are given by
\begin{align}
a_{m, \theta}=\int_{0}^{L}\frac{dy}{\sqrt{|E_{m,\theta}|}}e^{-\frac{i\theta y}{2s}} [u_{m,\theta}\partial_{y}\phi_{1}(y)+w_{m,\theta}\partial_{y}\phi_{2}(y)].
\label{amtheta}
\end{align}
where $u_{m,\theta}$ and $w_{m,\theta}$ are periodic functions of $y$ to be derived below. Likewise, we find that the corresponding energies $E_{m,\theta}$ are given by (Fig.~\ref{fig:phonon_spectrum})
\begin{align}
E_{m,\theta}=\frac{(2m+1)\pi v}{4s} +  \frac{(-1)^m v}{s} \left[\arccos\left(\kappa\cos \frac{\theta}{2}\right) -\frac{\pi}{4}  \right].
\label{eq:spectrum}
\end{align}
where $\kappa \equiv \left|\frac{k_{1}-k_{2}}{k_{1}+k_{2}} \right|$.
Here the $a_{m,\theta}$ are normalized so that they obey the canonical commutation relations 
\begin{align}
[a_{m,\theta},a_{m,\theta}^\dagger]=\delta_{\theta, \theta'} \delta_{m m'}, \quad E_{m,\theta} > 0
\label{eq:comma}
\end{align}

With these results, we can immediately write down the low energy effective Hamiltonian in the limit $U \rightarrow \infty$:
\begin{align}
	H_\text{eff}=\sum_{m, \theta} \Theta(E_{m,\theta})E_{m,\theta} a^{\dagger}_{m,\theta}a_{m,\theta}
	\label{eq:Heff}
\end{align}
where $\Theta$ is the Heaviside step function. The main result of this analysis is that the phonon bands $E_{m, \theta}$ have a \emph{gap} around $E = 0$ (see Fig.~\ref{fig:phonon_spectrum}). From Eq.~(\ref{eq:spectrum}), we can see that the size of this gap is $\Delta_{\text{ph}} = \frac{v}{s} \arccos(\kappa)$.

\begin{figure}[t]
  \centering
\includegraphics[width=0.85\columnwidth]{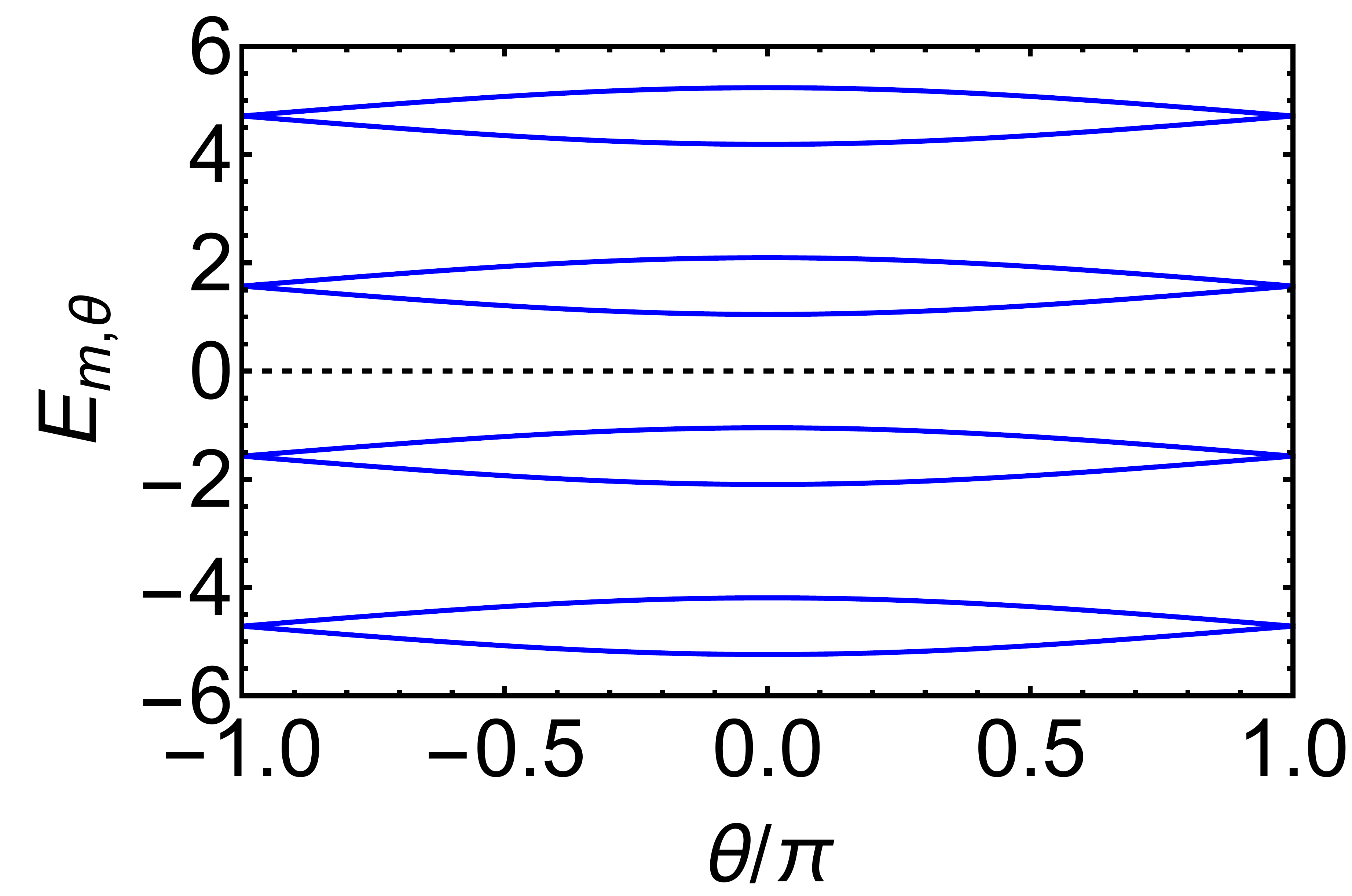}
       \caption{ Phonon spectrum (Eq.~[\ref{eq:spectrum}]) of the impurity model for the parameters $v=s=1$ and $(k_1, k_2) = (1,3)$. Notice the gap around $E=0$.}
  \la{fig:phonon_spectrum}
\end{figure}

We now derive the above results. First, we substitute (\ref{genaexp}) into (\ref{eq:aphonon}), which gives the differential equations
\begin{align*}
f_{1}'(y)& =  -i\frac{E}{v}f_{1}(y)- k_{1}\sum^{2N}_{l=1}\lambda_l \delta(y-l s)\nonumber\\
f_{2}'(y) & =  i\frac{E}{v}f_{2}(y)-k_{2}\sum^{2N}_{l=1}(-1)^{l+1}\lambda_l\delta(y-l s)
\end{align*}
Solving these equations gives piecewise plane waves
\begin{align}
f_{1}(y) &= A_{l}e^{-i\frac{E}{v}(y-ls)} \nonumber \\
f_{2}(y) &= B_{l}e^{i\frac{E}{v}(y-ls)}, \quad ls < y < (l+1)s
\end{align}
where $A_{l}, B_{l}$ satisfy the matching conditions
\begin{align*}
A_{l}&=A_{l-1}e^{-iEs/v}-\lambda_{l}k_1 , \\
 B_{l}&=B_{l-1}e^{iEs/v}-(-1)^{l+1}\lambda_{l}k_2.
\end{align*}
Eliminating $\lambda_l$, we derive
\begin{equation}
\frac{(A_{l} - A_{l-1}e^{-iEs/v})}{k_1} =(-1)^{l+1} \frac{(B_{l} - B_{l-1}e^{iEs/v})}{k_2} 
\label{constraintABfm1}
\end{equation}
We still have to impose the constraint $[a,C_l] = 0$, which gives an additional matching condition for $A_l, B_l$. After regularization of the cosine terms (see App.~B of Ref.~\onlinecite{heinrich2017solvable}), this constraint gives
\begin{equation}
\frac{(A_{l} + A_{l-1}e^{-iEs/v})}{2} =(-1)^{l+1} \frac{(B_{l} + B_{l-1}e^{iEs/v}) }{2}
\label{constraintABfm2}
\end{equation}
To proceed further, we assume $k_1 \neq k_2$; we discuss the case $k_1 = k_2$ later. Using (\ref{constraintABfm1}), (\ref{constraintABfm2}), we solve for $A_l, B_l$ in terms of $A_{l-1}, B_{l-1}$:
\begin{align}
\bpm
A_{l} \\
B_{l}
\epm = \mathcal{T}_\pm.
\bpm
A_{l-1}\\
B_{l-1}
\epm
\end{align}
where $\mathcal{T}_\pm$ is defined by
\begin{align}
\mathcal{T}_\pm&=\frac{1}{(k_{2}-k_{1})}\bpm
(k_{1}+k_{2})e^{-iEs/v} &  \mp 2k_{1}e^{iEs/v}\\
\pm 2k_{2}e^{-iEs/v} & -(k_{1}+k_{2})e^{iEs/v}
\epm \nonumber
\end{align}
and where we use $\mathcal{T}_+$ when $l$ is odd and $\mathcal{T}_-$ when $l$ is even. 

The transfer matrix across a full unit cell is given by the product of $\mathcal{T}_-$ and $\mathcal{T}_+$:
\begin{align*}
\bpm
A_{2l+2}\\
B_{2l+2}
\epm
=\mathcal{T}_- \mathcal{T}_+ \bpm
A_{2l}\\
B_{2l}
\epm
\end{align*}

Next we use the Bloch condition (\ref{eq:bloch}) to deduce that 
\begin{align*}
\bpm
A_{2l}\\
B_{2l}
\epm
=e^{-il\theta}
\bpm
A \\
B
\epm
\end{align*}
where $(A, B)\equiv (A_0, B_0)$. Combining the above two equations, we obtain: 
\begin{align}
\mathcal{T}_- \mathcal{T}_+ \bpm A \\ B \epm
=e^{-i\theta} \bpm A \\ B \epm
\end{align}
The above eigenvalue equation completely determines the phonon dispersion. To solve this equation, notice that $\det(\mathcal{T}_+)=\det(\mathcal{T}_-)=-1$ so that $\det(\mathcal{T}_- \mathcal{T}_+)=1$. Thus if $\mathcal{T}_- \mathcal{T}_+$ has $e^{-i\theta}$ as an eigenvalue, then it has $e^{i\theta}$ as the other eigenvalue. Hence, $\text{Tr}(\mathcal{T}_-\mathcal{T}_+)=2\cos \theta$.
Writing this equation out explicitly gives
\begin{align*}
	\cos(\theta) & = \left(\frac{k_{1}+k_{2}}{k_{1}-k_{2}}\right)^{2}\cos\left(2Es/v\right)+\frac{4k_{1}k_{2}}{(k_{1}-k_{2})^{2}}
\end{align*}
Using a trigonometric identity, this can be further simplified to:
\begin{align*}
	\cos\left(Es/v\right) & =\pm\left(\frac{k_{1}-k_{2}}{k_{1}+k_{2}}\right)\cos(\theta/2)
\end{align*}
Notice that for each $\theta\in [-\pi, \pi]$, there are infinitely many $E$'s that satisfy the above equation. These are precisely the energies $E_{m,\theta}$ presented in Eq.~(\ref{eq:spectrum}).
A little algebra shows that the corresponding values for $A,B$ are given by
\begin{align}
A_{m,\theta} &= (k_{1}+k_{2})^2e^{\frac{-2isE_{m,\theta}}{v}}-(k_{1}-k_{2})^2e^{i \theta}+4k_1 k_2 \nonumber \\
B_{m,\theta} &= -(1+e^{\frac{-2isE_{m,\theta}}{v}})2k_2(k_1+k_2)
\end{align}
We can now write down the explicit form of the two functions $u_{m, \theta}$ and $w_{m, \theta}$ in Eq.~(\ref{amtheta}):
\begin{align}
	u_{m,\theta} &=\frac{A^\pm_{m,\theta}}{\mathcal{N}_{m,\theta}}e^{i(\theta/2s-E_{m,\theta}/v)(y - ls)} \\
	w_{m,\theta} &=\frac{B^\pm_{m,\theta}}{\mathcal{N}_{m,\theta}}e^{i(\theta/2s+E_{m,\theta}/v)(y - ls)}, \quad ls \leq y \leq (l+1)s 
\end{align}
where 
\begin{align*}
\bpm A^-_{m, \theta} \\ B^-_{m,\theta} \epm \equiv \bpm A_{m, \theta} \\ B_{m, \theta} \epm, \quad 
\bpm A^+_{m, \theta} \\ B^+_{m,\theta} \epm \equiv e^{i \theta/2}\mathcal{T}_+ \bpm A_{m, \theta} \\ B_{m, \theta} \epm,
\end{align*} 
and where we use $A^-, B^-$ when $l$ is even and $A^+$ and $B^+$ when $l$ is odd. 

The normalization constant $\mathcal{N}_{m,\theta}$ is given by 
 \begin{align}
 	 \mathcal{N}_{m,\theta}=\sqrt{\frac{2\pi L}{v}}\left(\frac{|A_{m,\theta}|^2}{k_1}+\frac{|B_{m,\theta}|^2}{k_2}\right)^{1/2}
 \end{align}
and is obtained by demanding that $a_{m,\theta}$ obey the canonical commutation relations (\ref{eq:comma}). 

As we mentioned earlier, the above derivation assumes that $k_1 \neq k_2$ since the transfer matrices $\mathcal{T}_\pm$ are not well-defined when $k_1 = k_2$. Therefore, strictly speaking, we need a separate derivation for the case $k_1 = k_2$. However, the $k_1 = k_2$ case can be analyzed straightforwardly starting from Eqs.~(\ref{constraintABfm1}), (\ref{constraintABfm2}), as discussed in Ref.~\onlinecite{ganeshan2016formalism}. The end result for the energy spectrum is the same as one would get by naively substituting $k_1 = k_2$ into (\ref{eq:spectrum}). Thus, the above results hold for both $k_1 = k_2$ and $k_1 \neq k_2$.

\subsection{Ground state degeneracy}
\label{sec:gsd}
In the previous section, we showed that all the phonon modes of $H$ (\ref{Hfmsc}) are gapped in the limit $U \rightarrow \infty$. This means that the only possible low energy states of $H$ are its ground states. In this section, we compute this ground state degeneracy and show that it grows exponentially with $N$.

We will use the formalism of Sec.~\ref{sec:rof} for this calculation, but we first need to take care of a technical problem. The problem is that our formalism assumes that all degrees of freedom are continuous and real valued, but our system has two integer valued degrees of freedom, namely the total charge $Q_1, Q_2$ (\ref{Qidef}) on the two edge modes. The discrete nature of $Q_1, Q_2$ is important for obtaining the correct ground state degeneracy so we need to take account of it in the remainder of our analysis. We do this using a trick: we treat all degrees of freedom as though they are real valued and then we introduce two additional cosine terms to enforce the discreteness of $Q_1, Q_2$ at an energetic level, namely 
\begin{align*}
H \rightarrow H - U \cos(2\pi Q_1) - U \cos(2\pi Q_2)
\end{align*}
In the limit $U \rightarrow \infty$, these terms lock $Q_1, Q_2$ to integer values and also make the corresponding conjugate variables $\phi_1, \phi_2$ compact.
With this trick, we now have $2N+2$ cosine terms $\cos(C_l)$ with
\begin{align*}
&C_l = (k_{1}\phi_{1}(l s)+(-1)^{l+1}k_{2}\phi_{2}(l s))\quad l=1,...,2N \\
&C_{2N+1} = 2\pi Q_1, \quad C_{2N+2}= 2\pi Q_2.
\end{align*}

We are now ready to compute the ground state degeneracy using the general formalism. The first step is to compute the commutator matrix $\mathcal{Z}_{ij} = \frac{1}{2\pi i} [C_i,C_j]$. This can be done straightforwardly once we choose a convention for the commutation relations for $\phi_1, \phi_2$. We use the following convention:
\begin{align*}
[\phi_1(l s), \phi_1(l' s)] &= -\frac{\pi i}{k_1} \text{sgn}(l'-l) \\
[\phi_2(l s), \phi_2(l' s)] &= \frac{\pi i}{k_2} \text{sgn}(l'-l) 
\end{align*}
With this convention, we find
\begin{eqnarray}
\hspace{-10mm}\mathcal{Z}_{ij} 
&=& \bpm 0 & -\alpha & \beta & -\alpha & \cdots & \beta & -\alpha & -1 & 1 \\ 
	\alpha & 0 & -\alpha & \beta & \cdots &-\alpha & \beta & -1 & -1 \\
	 -\beta &\alpha & 0 & -\alpha & \cdots & \beta & -\alpha & -1 & 1 \\
	\alpha & -\beta &\alpha & 0 & \cdots & -\alpha & \beta & -1 & -1 \\
	\vdots & \vdots & \vdots & \vdots & \vdots & \vdots & \vdots & \vdots & \vdots \\
	 -\beta & \alpha & -\beta & \alpha & \cdots & 0 & -\alpha & -1 & 1 \\
	\alpha & -\beta & \alpha & -\beta & \cdots & \alpha & 0 & -1 & -1 \\
	1 &1 & 1 & 1 & \cdots & 1 &1 & 0 & 0 \\
	-1 & 1 &  -1 & 1 & \cdots & -1 & 1 & 0 & 0 \epm
	\label{zij}
\end{eqnarray}
where 
\begin{align}
\alpha=\frac{k_1+k_2}{2}, \quad \beta=\frac{k_2-k_1}{2}.
\end{align}


The next step is to compute the ground state degeneracy using the formula (\ref{degformula}). According to this formula, the ground state degeneracy is given by $\sqrt{|\det{\mathcal{Z}}|}$. We denote this quantity by $D_N$ (for a system of $2N$ impurities). $D_N$ can be determined by simplifying the $\mathcal{Z}_{ij}$ matrix using row and column operations, $R_i\rightarrow R_i-R_{i+2}$  and $C_i\rightarrow C_i-C_{i+2}$ for $i=1,...,2N-2$. The transformed matrix is of the form
\begin{eqnarray*}
\bpm 0 & -2\alpha & \beta & 0 & \cdots & 0 & 0 & 0 & 0 \\ 
	2\alpha & 0 & -2\alpha & \beta & \cdots & 0 & 0 & 0 & 0 \\
	 -\beta &2\alpha & 0 & -2\alpha & \cdots & 0 & 0 & 0 & 0 \\
	0 & -\beta &2\alpha & 0 & \cdots & 0 & 0 & 0 & 0 \\
	\vdots & \vdots & \vdots & \vdots & \vdots & \vdots & \vdots & \vdots & \vdots \\
	 0 & 0 & 0 & 0 & \cdots & 0 & -\alpha & -1 & 1 \\
	0 & 0 & 0 & 0 & \cdots & \alpha & 0 & -1 & -1 \\
	0 & 0 & 0 & 0 & \cdots &1 & 1 & 0 & 0 \\
	0 & 0 &  0 & 0 & \cdots & -1 & 1 & 0 & 0 \epm,
\end{eqnarray*}
Computing $D_N$ using the standard recursive formula for the Pfaffian of a skew-symmetric matrix, we derive the following recursion relation:
\begin{align}
D_N &=2 \alpha D_{N-1}-\beta^2 D_{N-2} 
\label{eq:rec}	
\end{align}
Solving the recursion relation with $D_1 = 2$ and $D_2 = 4\alpha$, gives the following explicit formula for $D_N$:
\begin{align}
D_N=\frac{\left(\frac{k_1+k_2}{2}+\sqrt{k_1k_2}\right)^{N}-\left(\frac{k_1+k_2}{2}-\sqrt{k_1k_2}\right)^N}{\sqrt{k_1k_2}}.
\label{dnform}
\end{align}	
In the limit of large $N$, we see that the degeneracy $D_N$ grows exponentially with $N$ as
\begin{align}
D_N\sim\frac{1}{\sqrt{k_1k_2}}\left(\frac{k_1+k_2}{2}+\sqrt{k_1k_2}\right)^{N}
\end{align}

\begin{figure}[t]
  \centering
\includegraphics[width=0.6\columnwidth]{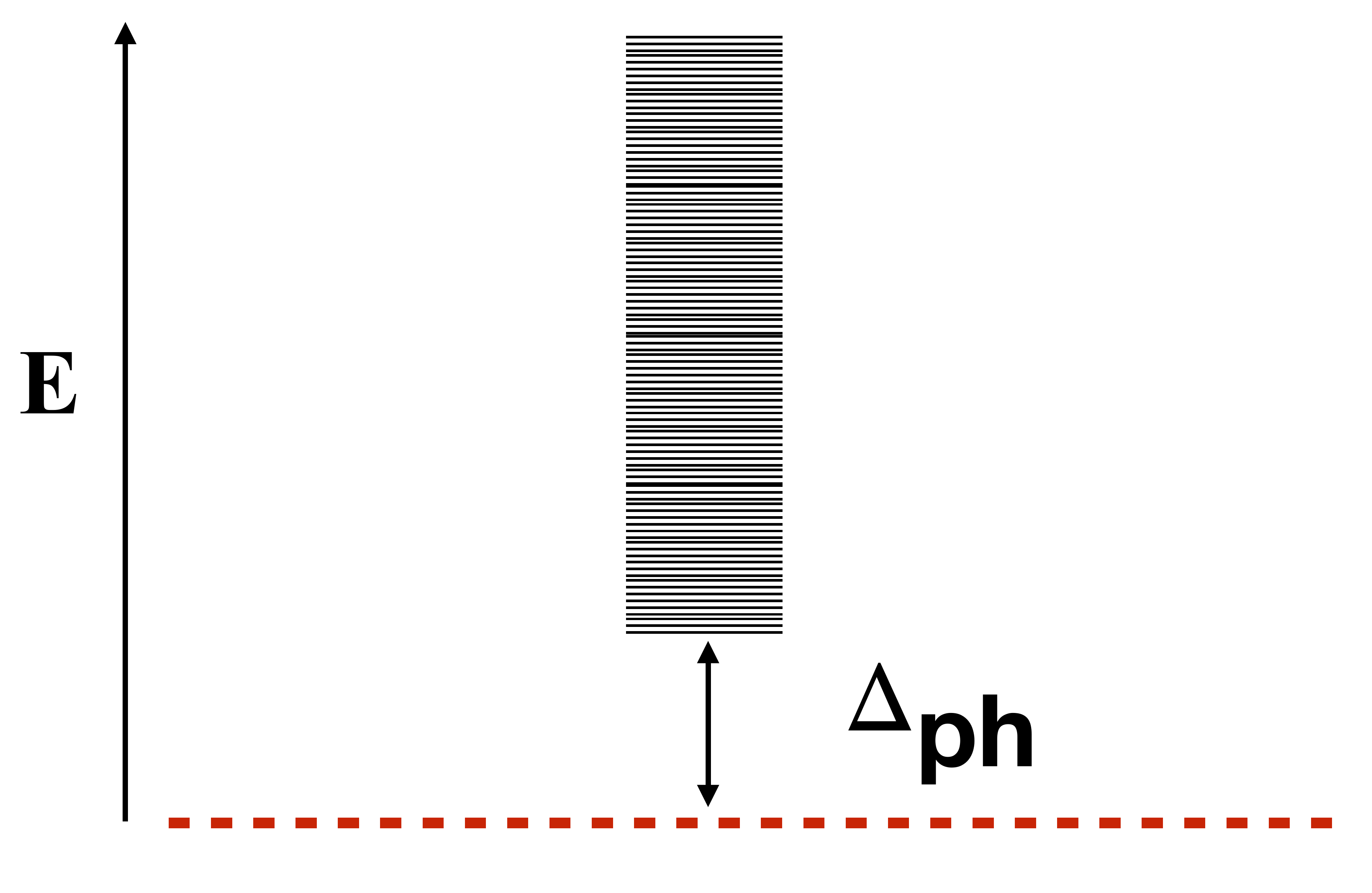}
    \caption{Many-body energy spectrum of impurity model in $U \rightarrow \infty$ limit: gapped phonon modes and an extensive ground state degeneracy. }
    \label{fig:spsch}
\end{figure}

To summarize, we have shown that the impurity model $H$ (\ref{Hfmsc}) has two properties in the limit $U \rightarrow \infty$: (1) the phonon modes of $H$ have a gap of size $\Delta_{\text{ph}} = \frac{v}{s} \arccos(\kappa)$ (see Eq.~[\ref{eq:spectrum}]) and (2) $H$ has a ground state degeneracy $D_N$ (\ref{dnform}). Together, these two results imply that the energy spectrum of $H$ looks like the one shown in Fig.~\ref{fig:spsch}.

\section{Finite dimensional edge theory}
\label{sec:edgeth}

We now derive our edge theory by projecting into the ground state subspace of $H$ (Fig.~\ref{fig:spsch}).




\subsection{Deriving the edge theory}
To construct our edge theory we need to specify two pieces of data: (1) the Hilbert space $\mathcal{H}$ and (2) the set of local operators $\{\mathcal{O}\}$ in our edge theory. 

We start with the Hilbert space $\mathcal{H}$. In our case, $\mathcal{H}$ is simply the ground state subspace of the impurity model. This subspace consists of all states $|\psi\>$ obeying two conditions:
\begin{align}
a_{m,\theta}|\psi\> = 0, \quad \quad \cos(C_j) |\psi\> = |\psi\>.
\label{Hilbdef}
\end{align}
Here, the $\{a_{m,\theta}\}$ are the phonon annihilation operators (\ref{amtheta}) while the $\{C_j\}$ are given in (\ref{eq:cons}). 

Next we discuss the local operators $\{\mathcal{O}\}$ in our edge theory. These operators are defined by projecting the local operators in the original chiral boson edge theory into the ground state subspace (\ref{Hilbdef}). To analyze this projection, we use the formalism of Ref.~\onlinecite{ganeshan2016formalism}. First, we note that according to App.~D3 of Ref.~\onlinecite{ganeshan2016formalism}, the most general ground state operator can be written as a polynomial in $e^{\pm i \Gamma_i}$ where 
\begin{align}
\Gamma_i = \sum^{2N+2}_{j=1} \mathcal{Z}^{-1}_{ji} C_j.
\label{Gammadef0}
\end{align}
At the same time, it is clear from physical considerations that $e^{i \Gamma_i}$ corresponds to a \emph{local} operator near impurity $i$ (see below for an explanation). Combining these two facts, we conclude that $\{e^{ i \Gamma_1}, ..., e^{ i \Gamma_{2N}}\}$ are the basic local operators in the edge theory. Equivalently, the basic local operators are $\{U_1,...,U_{2N}\}$ where
\begin{align}
U_i= e^{i \Gamma_i + i \phi},
\label{Gammadef}
\end{align}
and where $\phi$ is an additional phase that we include in the definition of $U_i$ in order to simplify some of the equations below. Our specific choice of $\phi$ is given in Eq.~(\ref{phidef}). 

The $U_i$'s should be thought of as analogs of the Pauli spin operators $\{\sigma_i^x, \sigma_i^y, \sigma_i^z\}$ in a quantum spin-$1/2$ chain: any observable in the edge theory can be constructed by taking sums and products (and adjoints) of the $U_i$ operators. Likewise, any \emph{local} operator in the edge theory can be built out of $U_i$'s with $i$ restricted to a finite interval.

To understand the physical interpretation of the $U_i$ operators, notice that 
\begin{align}
[a_{m, \theta}, \Gamma_i] = 0, \quad \quad [C_j, \Gamma_i] = 2\pi i \delta_{ij}
\end{align}
where the first equality follows from (\ref{eq:cons}). These commutation relations have several implications. First, they imply that $U_i$ commutes with both $a_{m, \theta}$ and $\cos(C_j)$. This means that $U_i$ maps the ground state subspace to itself -- an important consistency check. A second implication is that $U_i^{-1} C_j U_i = C_j - 2\pi \delta_{ij}$. This identity gives a physical interpretation to $U_i$: we can think of $U_i$ as describing an instanton tunneling event where $C_i$ shifts by $2\pi$ (see App.~C of Ref.~\onlinecite{ganeshan2016formalism} for more details). This identity also explains why $U_i$ should be thought of as an operator that is localized near impurity $i$, as we claimed earlier.

One aspect of the $U_i$'s that is worth emphasizing is that they have \emph{even} fermion parity. This means that the $U_i$'s are not capable of describing fermion parity non-conserving processes like an electron tunneling into an edge from another system. To describe such processes, one needs to supplement the $U_i$'s with fermion parity-\emph{odd} operators. We construct these ``electron-like'' operators in Appendix~\ref{electronapp}; we will not need them in the main text.

\subsection{Algebraic definition of the edge theory}
\label{sec:alg_edge_theory}
In principle, our edge theory is fully defined by the Hilbert space (\ref{Hilbdef}) together with the local operators (\ref{Gammadef}). However, it is more convenient to define the edge theory in terms of the operator algebra obeyed by the $U_i$ operators; this is analogous to defining a spin-$1/2$ chain by the algebra of the Pauli spin operators. 

To this end, we now list the fundamental algebraic properties of the $U_i$ operators (see App.~\ref{derivrelapp} for a derivation). First,
\begin{align}
&U_i^{-1} = U_i^\dagger \label{alg0} \\
&U_i U_j =e^{2\pi i \mathcal{Z}^{-1}_{ij}} U_j U_i \label{alg1} \\
&U_i^{\beta} U_{i+1}^{-2\alpha}U_{i+2}^{\beta}=\mathbb{I} \label{alg2}
\end{align}
where the indices $i+1, i+2$ are defined \emph{modulo} $2N$. Also,
\begin{align}
U_1 U_2 \cdots U_{2N} &= e^{i \theta_1} \mathbb{I}, \nonumber \\
U_1^{-1} U_2 U_3^{-1} \cdots U_{2N} &=e^{i\theta_2} \mathbb{I} \label{alg3}
\end{align}
where (\ref{alg3}) can be thought of as global boundary conditions, and $\theta_1, \theta_2$ are two phases that specify these boundary conditions. In our setup, the values of $\theta_1, \theta_2$ are given by Eq.~(\ref{thetadef}). Finally, the $U_i$ operators obey a technical condition related to fermion parity: 
\begin{align}
\text{Tr}\left(U_2 U_4 \cdots U_{2N}\right) = 0
\label{alg4}
\end{align}
To understand the connection between (\ref{alg4}) and fermion parity, note that $\prod_{i=1}^N U_{2i} \propto \exp(i\sum_{i=1}^{N} \Gamma_{2i})$. The latter operator can be identified with the fermion parity operator via the identity $\sum_{i=1}^{2N} \Gamma_{2i} = -\pi(Q_1 + Q_2)$. Thus, (\ref{alg4}) says that the fermion parity operator in our edge theory has a vanishing trace. Equivalently, it says that our edge theory contains both even and odd fermion parity sectors with the same dimensionality. 

An important property of the above algebraic relations (\ref{alg0}-\ref{alg4}) is that they are \emph{complete}. In other words, there is a unique\footnote{We will not prove uniqueness in this paper.} representation of dimension $D_N$ of the algebra defined by Eqs.~(\ref{alg0}-\ref{alg4}). (See App.~\ref{sec:numerical} for an algorithm for constructing this representation). Thus, we can simply \emph{define} the edge theory Hilbert space $\mathcal{H}$ to be this unique representation. This algebraic definition is a compact way to describe the edge theory. 

We now discuss the structure of the $\mathcal{Z}^{-1}_{ij}$ matrix since it plays an important role in our edge theory. We focus on the $2N \times 2N$ \emph{submatrix} of $\mathcal{Z}^{-1}_{ij}$ corresponding to $i,j = 1,...,2N$, since this is what appears in the commutation relation (\ref{alg1}). It is not hard to show that this submatrix is given by
\begin{align}
(\mathcal{Z}^{-1})_{1 \leq i,j \leq 2N} = \mathcal{X}^{-1} \mathcal{S}
\label{eq:zij0}
\end{align}
where $\mathcal{X}, \mathcal{S}$ are the following $2N \times 2N$ matrices:
\begin{align}
\mathcal{X} &= \bpm 2\alpha & -\beta & 0 & \cdots & 0 & -\beta \\
					-\beta & 2\alpha & -\beta & \cdots & 0 & 0 \\
					0 & -\beta & 2\alpha & \cdots & 0 & 0 \\
					\vdots & \vdots & \vdots & \vdots & \vdots & \vdots \\
					0 & 0 & 0 & \cdots & 2\alpha & -\beta \\
					-\beta & 0 & 0 & \cdots & -\beta & 2\alpha \epm \label{Xdef} \\
\mathcal{S} &= \bpm 0 & 1 & 0 & \cdots & 0 & -1 \\
					-1 & 0 & 1 & \cdots & 0 & 0 \\
					0 & -1 & 0 & \cdots & 0 & 0 \\
					\vdots & \vdots & \vdots & \vdots & \vdots & \vdots \\
					0 & 0 & 0 & \cdots & 0 & 1 \\
					1 & 0 & 0 & \cdots & -1 & 0 \epm
\end{align} 
(This expression can be derived straightforwardly by performing a sequence of elementary row operations on $\mathcal{Z}$ similar to the ones discussed in Sec.~\ref{sec:gsd}). 

Eq.~(\ref{eq:zij0}) reveals several important properties of the $2N \times 2N$ submatrix of $\mathcal{Z}^{-1}_{ij}$. First, we can see that $\mathcal{Z}^{-1}_{ij}$ is translationally invariant since $\mathcal{X}, \mathcal{S}$ are translationally invariant. That is,
\begin{align}
\mathcal{Z}^{-1}_{(i+\ell)(j+\ell)}=\mathcal{Z}^{-1}_{ij}, \la{eq:transinv}
\end{align} 
where the sums $i+\ell$ and $j+\ell$ are defined modulo $2N$. The other important property of $\mathcal{Z}^{-1}_{ij}$ is that it is \emph{quasi-diagonal} in the sense that $\mathcal{Z}^{-1}_{ij}$ decays exponentially with increasing $|i-j|$. (This property follows from the fact $\mathcal{X}$ has eigenvalues that are bounded away from $0$). 
Note that this quasi-diagonal structure means that $U_i$ and $U_j$ approximately commute at large separations. 

A final application of (\ref{eq:zij0}) is that it can be used to derive a closed form expression for $\mathcal{Z}^{-1}_{ij}$ for $i,j=1,...,2N$. To do this, we use the following formula for the inverse of $\mathcal{X}$: 
\begin{align}
\mathcal{X}^{-1}_{ij} = \frac{2}{k_2 - k_1}\frac{x^{N-|i-j|} + x^{|i-j|-N}}{(x^N - x^{-N})(x-x^{-1})} 
\label{xinv}
\end{align}
where 
\begin{align}
x =\frac{k_1+k_2+2\sqrt{k_1k_2}}{k_2-k_1}
\label{xform}
\end{align}
Combining (\ref{eq:zij0}) and (\ref{xinv}) gives a closed form expression for $\mathcal{Z}^{-1}_{ij}$ for $i,j=1,...,2N$:
\begin{align}
\mathcal{Z}^{-1}_{ij}&=\text{sgn}(j-i)\frac{2}{k_2-k_1}\frac{x^{N-|i-j|}-x^{-N+|i-j|}}{x^{N}-x^{-N}},
\label{eq:zij}
\end{align}
where we define $\text{sgn}(0)=0$. This expression simplifies in the thermodynamic limit, $N\rightarrow \infty$: 
\begin{align}
\lim_{N \rightarrow \infty} \mathcal{Z}^{-1}_{ij} = 2\frac{x^{-|i-j|}}{k_2-k_1}\text{sgn}(j-i).
\label{Zapprox}
\end{align}
These formulas are useful because they make the algebraic relations for the $U_i$ operators more explicit, particularly Eq.~(\ref{alg1}).

Before concluding this section, we make two more comments. First, we note that the edge theory defined by (\ref{alg0}-\ref{alg4}) has a lattice translational symmetry. In particular, there exists a unitary translation operator $T$ that shifts each $U_i \rightarrow U_{i+2}$:
\begin{align}
T U_i T^{-1} = U_{i+2}
\end{align}
where $i+2$ is defined modulo $2N$. To see this, note that all of the defining relations (\ref{alg0}-\ref{alg4}) are invariant under replacing $U_i \rightarrow U_{i+2}$, since $\mathcal{Z}^{-1}$ is translationally invariant (\ref{eq:transinv}). This translational symmetry is physically reasonable since it matches the translational symmetry in our original impurity model (\ref{Hfmsc}) with alternating impurity types. 

Our second comment is about the special case $k_1 = k_2 = k$. In that case, the edge theory simplifies substantially. The first simplification is that $\mathcal{Z}^{-1}_{ij}$ becomes a tridiagonal matrix with $0$ on the main diagonal and $\pm \frac{1}{2k}$ on the neighboring diagonals. Eq.~(\ref{alg1}) then reduces to
\begin{align*}
U_i U_{i+1} = e^{i\pi/k} U_{i+1} U_i
\end{align*}
with $[U_i, U_j] = 0$ for $|i-j| \geq 1$. Also, since $\beta = 0$, Eq.~(\ref{alg2}) reduces to
\begin{align*}
U_i^{2k} = \mathbb{I}
\end{align*}
Not suprisingly, this operator algebra matches the one found in Refs.~\onlinecite{lindner2012fractionalizing,  cheng2012superconducting, clarke2013exotic, barkeshli2013twist}, which studied a closely related model consisting of a fractional quantum spin Hall edge proximity coupled to an alternating sequence of superconductors and ferromagnets. Here, the $U_i$ operators correspond to the charge and spin operators, $e^{i \pi Q_j}$ and $e^{i \pi S_j}$, discussed e.g.~in Ref.~\onlinecite{lindner2012fractionalizing}. 

\subsection{Anyonic string operators}
\label{sec:anyonstring}
We now discuss the ``anyonic string operators'' in our edge theory. Recall that an important aspect of edge theories of non-trivial topological phases is that they support non-local string operators which are parameterized by two endpoints $a$ and $b$ on the edge. These string operators have a simple physical interpretation: they describe processes in which a pair of anyons $\alpha, \bar{\alpha}$ are created in the bulk and then moved near the edge, where they are absorbed near two points $a, b$. At an algebraic level, these string operators have two crucial properties: (1) they commute with all local operators $\mathcal{O}$ except for those supported near their endpoints $a$ and $b$; and (2) they are \emph{nonlocal} in the sense that they cannot be written as products of local operators supported near the two endpoints $a, b$. 

We will argue below that the two basic string operators in our edge theory are given by
  	\begin{align}
		W^1_{ab}=\prod^b_{i=a}U_i,\quad 
		W^2_{ab}=\prod^b_{i=a}U_i^{(-1)^{i}}.
	\end{align}
where $a, b$ denote the endpoints of the string operators. Other string operators can be obtained by considering products of the form $(W^1_{ab})^{m_1} (W^2_{ab})^{m_2}$ with $0 \leq m_1 \leq k_1 -1$, and $0 \leq m_2 \leq k_2 - 1$. All together, this gives $k_1 k_2$ different string operators -- one for every anyon type.

To see that $W^1_{ab}$ and $W^2_{ab}$ are legitimate string operators, we first need to verify property (1), i.e. we need to check that $W^1_{ab}$ and $W^2_{ab}$ commute with all local operators $U_c$ except when $c$ is near $a, b$. To see this, we use the commutation algebra (\ref{alg1}) to derive
	\begin{align}
	W^{1}_{ab} U_c&= e^{i \vartheta_1}U_cW^{1}_{ab}, \quad \vartheta_1= 2\pi \sum^b_{i=a} \mathcal{Z}^{-1}_{ic} \\
W^{2}_{ab} U_c&=e^{i \vartheta_2}U_cW^{2}_{ab}, \quad \vartheta_2= \sum^b_{i=a}(-1)^i\mathcal{Z}^{-1}_{ic}
	\end{align}
There are two cases to consider: (i) $c$ could be outside the interval $[a, b]$, i.e. $a < b < c$, or (ii) $c$ could be inside $[a, b]$, i.e. $a < c < b$. In case (i), we can use the large $N$ limit of $\mathcal{Z}^{-1}_{ij}$ given in Eq.~(\ref{Zapprox}) to deduce that $\vartheta_1, \vartheta_2$ decay exponentially with the distance between $c$ and the interval $[a, b]$. In case (ii), we can use the global boundary conditions (\ref{alg3}) to rewrite $W^{1}_{ab}$ and $W^{2}_{ab}$ as products of $U_i$ over the \emph{complement} of $[a,b]$. Therefore, in this case, $\vartheta_1$ and $\vartheta_2$ decay exponentially with the distance between $c$ and the \emph{complement} of the interval $[a, b]$. Combining these two cases, we deduce that
\begin{align}
	\vartheta_i = \mathcal{O}(x^{-\min(|c-a|,|c-b|)}), \quad i = 1,2
\end{align}
This establishes property (1): $W^1_{ab}$ and $W^2_{ab}$ commute with all local operators $U_c$ except for those supported near $a, b$, up to an exponentially small error.

\begin{figure}[t]
  \centering
\includegraphics[width=0.5\columnwidth]{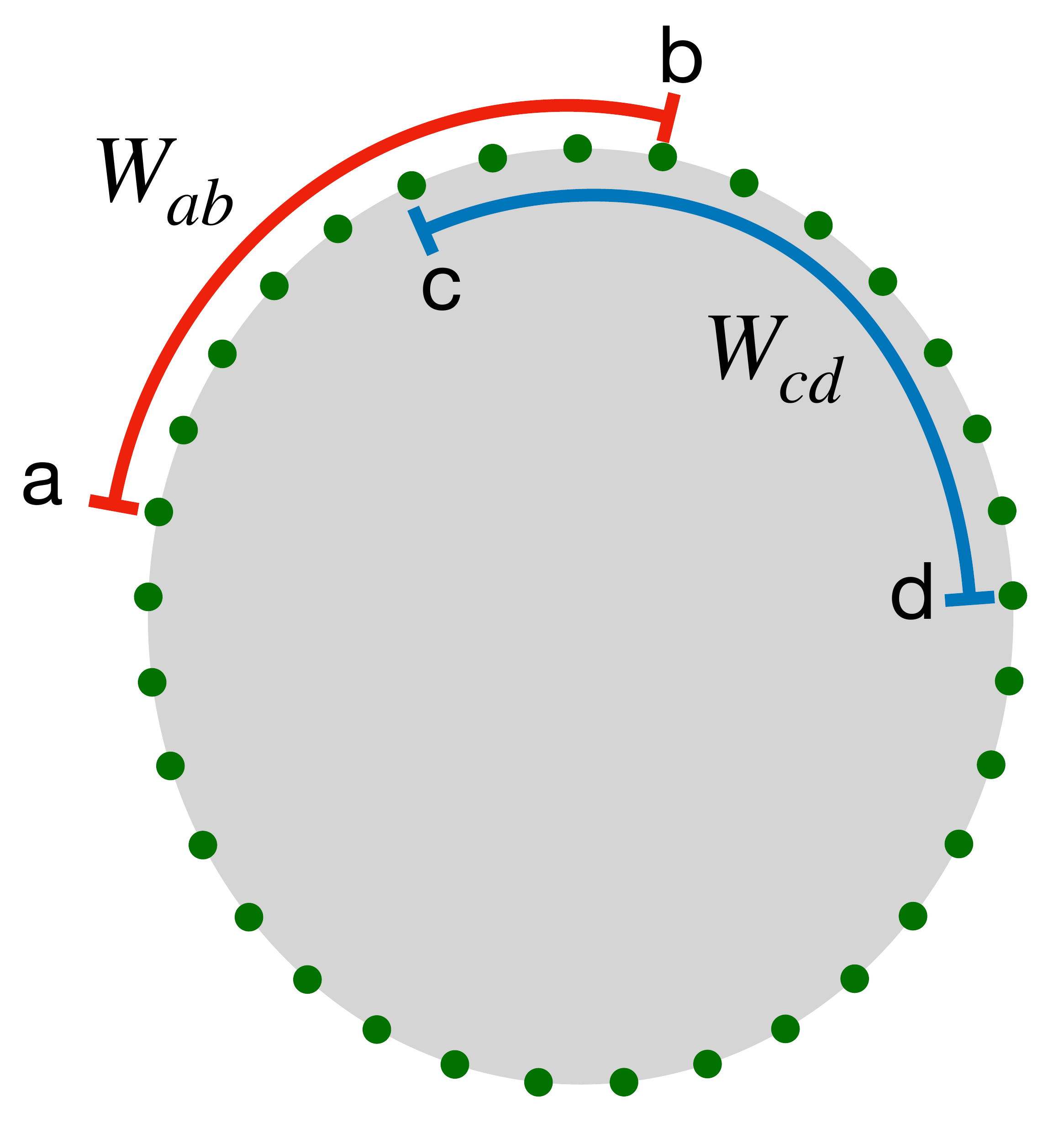}
    \caption{Two partially overlapping string operators, $W_{ab}, W_{cd}$.}
\la{fig:nonlocal2}
\end{figure}

Next we need to verify property (2), i.e. we need to check that $W^1_{ab}$ and $W^2_{ab}$ are nonlocal in the sense that they cannot be written as products of local operators supported near $a$ and $b$. The easiest way to establish this fact is to note that these string operators obey a nontrivial commutation algebra. In particular, consider two string operators $W^i_{ab}$ and $W^j_{cd}$ in an ``interleaved'' geometry with $a < c < b < d$ as in Fig.~\ref{fig:nonlocal2}. In this case, one finds that (see App.~\ref{sec:nonlocal})
\begin{align}
	W^i_{ab} W^j_{cd}&= e^{i\alpha_{ij}}W^j_{cd} W^i_{ab},
	\label{eq:nonlocal1}
\end{align}
where $\alpha_{ij}$ with $i,j = 1,2$, is the $2 \times 2$ matrix
\begin{align}
\alpha_{ij} = \bpm \frac{2\pi}{k_1} & 0 \\ 0 & -\frac{2\pi} {k_2} \epm + \mathcal{O}(x^{-\min(|a-c|, |b-c|, |a-d|, |b-d|)})
\end{align}
In other words, each pair of operators $(W^i_{ab}, W^i_{cd})$ obey a commutation algebra with a phase factor of $e^{\pm i 2\pi/k_i}$, up to exponentially small error. This nontrivial algebra implies that $W^i_{ab}$ cannot be written (or approximated) by a product of local operators supported near $a$ and $b$, thus proving property (2).

Before concluding this section, we should mention that the $W^i_{ab}$ operators are closely related to the anyonic string operators for the standard chiral boson edge theory, namely
\begin{align*}
\overline{W}^i_{ab} = \exp\left( i \int_a^b \partial_x \phi_i dx\right), \quad i = 1,2
\end{align*}
In particular, it is easy to check that the commutation algebra for the $W^i_{ab}$ operators is identical to that of $\overline{W}^i_{ab}$. This matching suggests that the $W^i_{ab}$ and $\overline{W}^i_{ab}$ operators correspond to the same bulk anyons, since the commutation algebra of anyonic string operators is directly related to the braiding statistics of the corresponding anyons. Indeed, using the formalism of Ref.~\onlinecite{ganeshan2016formalism}, one can check that \emph{projecting} the chiral boson string operator $\overline{W}^i_{ab}$ into the ground state subspace of the impurity model gives precisely the string operator $W^i_{ab}$.

\section{Gapping the $(1,9)$ edge theory: A type-I example}	
\label{sec:gap89}

As we mentioned in the introduction, the above family of edge theories describes both type-I and type-II topological phases: the type-I edge theories are those for which $\sqrt{k_1 k_2}$ is an integer, while the other theories are type-II. The main physical difference between the two types of edge theories involves the question of ``gappability'', i.e. whether there exist edge Hamiltonians that (i) are local in the sense that they can be written as a sum of local operators $\{\mathcal{O}\}$, and that (ii) have a gapped spectrum and a unique ground state. We expect that such gapping Hamiltonians can be constructed for type-I edge theories but not for type-II edge theories.

In this section, we verify this expectation for one of the simplest type-I edge theories, namely $(k_1, k_2) = (1,9)$. This edge theory can be thought of as describing the boundary of the $\nu = 8/9$ FQH state in a scenario where charge conservation symmetry is broken.  Our main result is the construction of a concrete gapping Hamiltonian for this edge theory. 


Our gapping Hamiltonian is given by
\begin{align}
H_{\text{gap}}=-\frac{1}{2}\sum_{i=1}^{2N}(K_i+h.c.)  -\frac{1}{2} \sum_{i=1}^{N}(L_i+h.c.)
\label{eq:hgap}
\end{align}
where $K_i, L_i$ are two site and three site operators of the form
\begin{align}
K_i&= e^{i \varphi_K} U_i^4 U_{i+1}^{-2}, \nonumber \\
L_i &= e^{i \varphi_L} U_{2i-1}^2 U_{2i}^{-5} U_{2i+1}^2 \label{gapterms}
\end{align}
Here $\varphi_K$ and $\varphi_L$ are two phases whose specific values are given in (\ref{phikdef}) and (\ref{phildef}). These values will not be important for our analysis. Also, as always, the indices for the $U_i$ operators are defined modulo $2N$. (See Sec.~\ref{sec:abscomm} for an explanation of how we found the Hamiltonian $H_{\text{gap}}$).

A crucial property of the $K_i$ and $L_i$ operators is that they commute with one another:
\begin{align}
[K_i, K_j] = [K_i, L_j] = [L_i, L_j] = 0 \label{commid}
\end{align}
In addition the $K_i$ and $L_i$ operators obey the following identities, for a suitable choice of phases $\varphi_K, \varphi_L$:
\begin{align}
K_i^2&=K_{i-1}, \label{shiftid} \\
K_{2N}^{Q}&= \mathbb{I}, \label{Qid}\\
L_i^2&=\mathbb{I} \label{Liid}	
\end{align}
where $Q = \frac{1}{3}(2^{2N}-1)$. We derive (\ref{commid})-(\ref{Liid}) in App.~\ref{derivklapp}. 

Using the above identities, we now proceed to compute the energy spectrum of $H_{\text{gap}}$. To this end, let us consider the collection of (commuting) operators $\{K_{2N}, L_1,...,L_N\}$. From (\ref{Qid}), we know that the eigenvalues of $K_{2N}$ are $Q$th roots of unity. Likewise, from (\ref{Liid}) we know that the eigenvalues of $L_i$ belong to the set $\{\pm 1\}$. Therefore, we can label the simultaneous eigenstates of $\{K_{2N}, L_1,...,L_N\}$ as $|m ; \sigma_1,..., \sigma_N\>$ where $m = 0,1,...,Q-1$ describes the eigenvalue of $K_{2N}$ and $\sigma_i = \pm 1$ describes the eigenvalue of $L_i$. That is:
\begin{align}
	K_{2N}|m; \sigma_1,..., \sigma_N\> &= e^{i 2\pi m/Q}|m; \sigma_1,..., \sigma_N\>\nn\\
L_i |m; \sigma_1,...\> &= \sigma_i |m; \sigma_1,...,\sigma_N\>\nn
\end{align}	
In view of (\ref{shiftid}), the states $|m ; \sigma_1,..., \sigma_N\>$ are also simultaneous eigenstates of $K_1,...,K_{2N-1}$, with eigenvalues
\begin{align}
K_{2N-n} |m; \sigma_1,..., \sigma_N\> = e^{i 2^n 2\pi m/Q}|m; \sigma_1,..., \sigma_N\>
\end{align}
Putting this all together, it follows that the $|m; \sigma_1,..., \sigma_N\>$ states are energy eigenstates of $H_{\text{gap}}$ with eigenvalue
\begin{align}
E(m, \{\sigma_i\}) = - \sum_{n=0}^{2N-1} \cos(2^n 2\pi m/Q) - \sum_{i=1}^N \sigma_i
	\label{eq:energy}
\end{align}
At this point, we \emph{almost} have the complete energy spectrum of $H_{\text{gap}}$. The only remaining issue is to determine the degeneracy of each simultaneous eigenspace labeled by $\{m, \sigma_1,...,\sigma_N\}$. We compute this degeneracy in App.~\ref{sec:deg}. We find that there is a \emph{unique} state for every choice of $\{m, \sigma_1,...,\sigma_N\}$. Note that this counting is consistent with the total dimension of the Hilbert space since there are $Q \cdot 2^{N} = \frac{2^{N}}{3}(2^{2N}-1)$ choices of the quantum numbers $m, \sigma_1,..., \sigma_N$, which exactly matches the dimension of the Hilbert space $D_N = (8^N-2^N)/3$, as given by (\ref{dnform}).

With  (\ref{eq:energy}) in hand, we can now read off the ground state and the energy gap. Specifically, we see that the ground state is the (unique) state where $m = 0$ and $\sigma_i = +1$ for all $i$. We can also see that there are two types of low energy excitations: ``spin flip'' excitations where $\sigma_i = -1$ for some $i$ and ``clock'' excitations where $m$ is nonzero. To find the energy gap we need to compute the energies of these two types of excitations. We start with the spin flip excitation, which has an energy of $\Delta_{\text{flip}} = 2$ since $\sigma_i = +1$ in the ground state and $\sigma_i = -1$ in the excited state. Moving on to the clock excitations, these have an energy of\footnote{Interestingly, $\Delta_{\text{clock}}(m)$ resembles a discretized version of the Weierstrass function $f(x) = \sum a^n \cos(b^n \pi x)$ -- a famous example of a function that is continuous everywhere but differentiable nowhere.}
\begin{align}
\Delta_{\text{clock}}(m) &= E(m, \{\sigma_i =1\})-E(0, \{\sigma_i=1\}) \nonumber \\
&= \sum_{n=0}^{2N-1}\left(1-\cos(2^n 2 \pi m/Q)\right)
\end{align}
It is easy to check numerically that the lowest energy clock excitation occurs at $m = 1$.\footnote{More generally, one can check that the lowest energy clock excitations form a degenerate multiplet of size $4N$, and occur at $m$'s of the form $m = \pm 2^k$ for $k=0,1,...,2N-1$.} Specializing to the lowest energy ($m=1$) excitation, we obtain
\begin{align*}
\Delta_{\text{clock}}(1) = \sum_{n=0}^{2N-1}\left(1-\cos(2^n 2\pi /Q)\right)
\end{align*}
Substituting $Q = \frac{1}{3}(2^{2N}-1)$ and taking the thermodynamic limit $N \rightarrow \infty$ gives
\begin{align}
\lim_{N \rightarrow \infty} \Delta_{\text{clock}}(1) &= \sum_{n=0}^{\infty}\left(1-\cos(3\pi 2^{-n})\right) \nonumber \\
&\approx 5.55
\end{align}
Combining these calculations, we conclude that the overall energy gap is
\begin{align}
\Delta = \text{min}(\Delta_{\text{flip}}, \Delta_{\text{clock}}(1)) = 2
\end{align}
In particular, we see that $H_{\text{gap}}$ has a finite energy gap in the thermodynamic limit, as we wished to show.

\section{Obstruction to gapping the $(1,3)$ edge theory: A type-II example}
\label{sec:obstg}
In this section, we investigate the gappability of one of the simplest type-II edge theories, namely $(k_1, k_2) = (1,3)$. This edge theory can be thought of as describing the boundary of the $\nu = 2/3$ FQH state in a scenario where charge conservation symmetry is broken. We consider two possible types of gapping Hamiltonians. The first type of Hamiltonian is a sum of \emph{commuting} operators (like the $H_{\text{gap}}$ Hamiltonian that we found in the $(k_1, k_2) = (1,9)$ case) while the second is a sum of \emph{non-commuting} operators. In both cases, we encounter obstructions to gapping the edge.

\subsection{Absence of local commuting operators}
\label{sec:abscomm}
We start by searching for local commuting operators for the $(k_1, k_2) = (1,3)$ edge theory. We focus on the simplest class of local operators, namely those of the form
\begin{align}
H_i = U_{i}^{a_0} U_{i+1}^{a_1} \cdots U_{i+m}^{a_{m}}
\label{Hidef}
\end{align}
where $a_0,...,a_m$ are integers. Such operators can be thought of as analogs of Pauli strings.

The main result of this section is that there are no non-trivial operators of the form (\ref{Hidef}) that commute with their translations, i.e.~no non-trivial operators that satisfy
\begin{align}
[H_i, H_j] = 0
\end{align}
More precisely, we show that if $[H_i, H_j] = 0$ for all $i,j$, and for arbitrarily large system sizes $N$, then $H_i = (const) \cdot \mathbb{I}$. This result means that the $(1,3)$ edge theory does \emph{not} support commuting Hamiltonians like the $H_{\text{gap}}$ Hamiltonian that we found in the $(1,9)$ case. 

The first step of the proof is to substitute the commutation relations for $U_i$ into $[H_i, H_j] = 0$. This yields the condition
\begin{align}
\sum_{k,l=0}^m a_k a_l \mathcal{Z}^{-1}_{(i+k)(j+l)} = 0 \pmod{2\pi}. 
\end{align}
Next we take the limit $N \rightarrow \infty$ and use the formula for $\mathcal{Z}^{-1}_{ij}$ (\ref{Zapprox}) to deduce
\begin{align}
\sum_{k,l=0}^m a_k a_l &\left( \frac{2x^{-|i+k-j-l|}}{k_2-k_1} \text{sgn}(j+l-i-k) \right) \nonumber \\
& \hspace{3.7cm}= 0\pmod{2\pi} 
\label{commform}
\end{align}
where $x$ is defined in (\ref{xform}). Note that we will ultimately specialize to the case $(k_1, k_2) = (1,3)$, but we keep these variables general for now. 

Next notice that the left hand side of (\ref{commform}) is exponentially small in the separation $|i-j|$. In particular, the left hand side is strictly greater than $-2\pi$ and strictly less than $2\pi$ for sufficiently large $|i-j|$. Hence, the equality must hold exactly, not just modulo $2\pi$. That is, 
\begin{align}
\sum_{k,l=0}^m a_k a_l \left( \frac{2x^{-|i+k-j-l|}}{k_2-k_1} \text{sgn}(i+k-j-l) \right)= 0 
\end{align}
for sufficiently large $|i-j|$. To proceed further, we specialize to the case where $i-j$ is large and \emph{positive}. In that case, the condition reduces to
\begin{align}
\sum_{k,l=0}^m a_k a_l \left( \frac{2x^{-(i+k-j-l)}}{k_2-k_1} \right)= 0 
\end{align}
Multiplying both sides by $\frac{(k_2-k_1) x^{i-j}}{2}$, we derive
\begin{align}
\sum_{k,l=0}^m a_k a_l x^{l-k} = 0
\end{align}
The left hand side can be factored as
\begin{align}
\left(\sum_{k=0}^m a_k x^{-k} \right) \left(\sum_{l=0}^m a_l x^l\right) = 0
\end{align}
so we deduce that one of the two terms on the left hand side must vanish. In other words, the polynomial
\begin{align}
P(z) \equiv \sum_{k=0}^m a_k z^k
\end{align}
has either $x$ or $x^{-1}$ as a zero. First, suppose that $x$ is a zero. In this case, we can use a standard theorem about algebraic numbers to deduce that $P(z)$ is divisible by $Q_+(z)$ where $Q_+(z)$ is the \emph{minimal} polynomial with integer coefficients that has $x$ as its zero. Likewise, if $x^{-1}$ is a zero then $P(z)$ must be divisible by $Q_-(z)$ where $Q_-(z)$ is the minimal polynomial that has $x^{-1}$ as its zero. 

Now we specialize to $(k_1, k_2) = (1, 3)$. Then $x = 2 + \sqrt{3}$ and $x^{-1} = 2 - \sqrt{3}$, and it is easy to see that the two minimal polynomials $Q_\pm(z)$ are equal and are given by
\begin{align}
Q_\pm(z) = z^2 -4 z + 1
\end{align}
Therefore, by the above argument, $P(z)$ must be divisible by $z^2 - 4z + 1$, i.e.
\begin{align}
P(z) &= \left(\sum_{l=0}^n b_l z^l \right) (z^2 - 4z + 1)
\end{align}
for some integers $b_l$. Equating coefficients of $z^{k}$ on the two sides gives the identity
\begin{align}
a_{k} = b_{k} - 4 b_{k-1} + b_{k-2}
\end{align}
(where we use the convention $b_{-1} = b_{-2} = 0$). The above identity in turn implies that
\begin{align}
H_i \propto O_i^{b_0} O_{i+1}^{b_1} \cdots O_{i+n}^{b_n}
\end{align}
where
\begin{align}
O_j = U_j U_{j+1}^{-4} U_{j+2}
\end{align}
The final step is to recall that $O_i = \mathbb{I}$ by the algebraic relation (\ref{alg2}) obeyed by the $U_i$ operators. Hence $H_i = (const) \mathbb{I}$, as we wished to show. As we mentioned earlier, this result rules out the possibility of constructing a commuting Hamiltonian for the $(1,3)$ edge theory like the one that we found in the $(1,9)$ case. 

The above analysis can also be extended to other choices of $k_1, k_2$. Perhaps the most interesting application is in the opposite direction -- i.e. finding gapping Hamiltonians for type-I edge theories, rather than finding obstructions to gapping type-II edge theories.  For example, consider the type-I edge theory with $(k_1, k_2) = (1, 9)$ discussed in Sec.~\ref{sec:gap89}. In this case, (\ref{xform}) gives $x=2$. Hence the minimal polynomials for $x$ and $x^{-1}$ are $Q_+(z) = z-2$, and $Q_-(z) = 2z-1$, respectively. Following the same logic as before, one deduces that the simplest candidates for commuting terms are $A_i = U_i^{-2} U_{i+1}$ (as well as $U_i U_{i+1}^{-2}$). By construction, the $A_i$ operators are guaranteed to commute approximately at large separations, but one can check that they actually commute \emph{exactly} at all separations -- except when they are nearest neighbors, in which case they anticommute. One can then construct fully commuting terms by considering the combinations $K_i \propto A_i^{-2}$ and $L_i \propto A_{2i-1}^{-1} A_{2i}^{2}$. Indeed, this line of reasoning is what led us to the gapping Hamiltonian in Sec.~\ref{sec:gap89}.

\subsection{Numerical study of non-commuting Hamiltonian}
\label{sec:numerics}
In this section we discuss the low energy spectrum of the following edge Hamiltonian, which is built out of \emph{non-commuting} local operators:
\begin{align}
H_{\text{nc}} =-\sum^{2N}_{i=1}(U_i+U_i^\dagger)\label{pert}
\end{align}
We consider $H_{\text{nc}}$ for two different type-II edge theories, namely $(k_1, k_2) = (1,3)$ and $(k_1, k_2) = (1,5)$. Our main result is that $H_{\text{nc}}$ is \emph{gapless} in both cases -- consistent with expectations. 

To obtain our results, we use numerical exact diagonalization. The numerical implementation is mostly straightforward; the only nontrivial step is the construction of an explicit matrix representation of $H_{\text{nc}}$, which we explain in App.~\ref{sec:numerical}.

\begin{figure}[t]
  \centering
  \includegraphics[width=0.85\columnwidth]{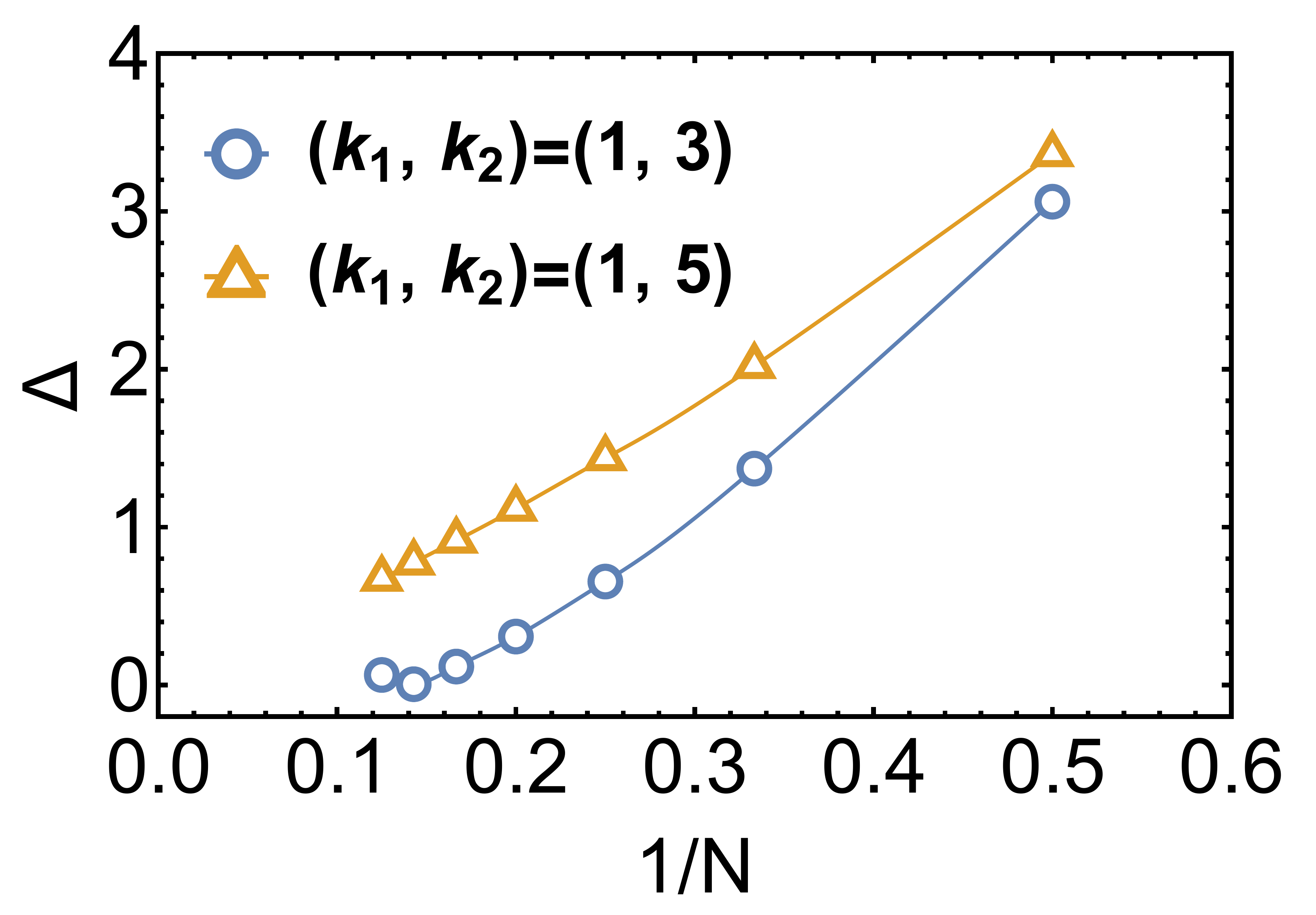}\
    \caption{Gap $\Delta$ between the ground state and first excited state of $H_{\text{nc}}$ (\ref{pert}) as a function of inverse size $1/N$ for the two cases $(k_1,k_2)=(1,3)$ and $(k_1,k_2)=(1,5)$.}
  \la{fig:gapscaling}
\end{figure}

To begin, we compute the gap $\Delta$ between the ground state and the first excited state as a function of the system size $N$. The results for $N = 2,3,...,8$ are shown in Fig.~\ref{fig:gapscaling}. For the $(1,5)$ case, our results are consistent with a gap that scales like $\Delta \propto 1/N$ for large $N$. For the $(1,3)$ case, we do not see simple scaling behavior at these system sizes, but our results are still consistent with a gap that vanishes in the thermodynamic limit.

Next, in order to get a more detailed picture of the low energy spectrum of $H_{\text{nc}}$, we use the fact that $H_{\text{nc}}$ is invariant under the discrete translational symmetry operator $T$ defined by $T U_i T^{-1} = U_{i+2}$ (see Sec.~\ref{sec:alg_edge_theory} for a discussion). This means that every energy eigenstate $|\Psi_n\>$ can be labeled by both its energy $\epsilon_n$ and its crystal momentum $\theta_n \in [0, 2\pi)$, where $\theta_n$ is defined by
\begin{align}
T |\Psi_n\> = e^{i \theta_n} |\Psi_n\>
\end{align}
Note that $\theta_n$ takes values in the set $\{0, 2\pi/N,...,2\pi(N-1)/N\}$ for a system with $2N$ impurities.

\begin{figure}[t]
  \centering
\includegraphics[width=0.9\columnwidth]{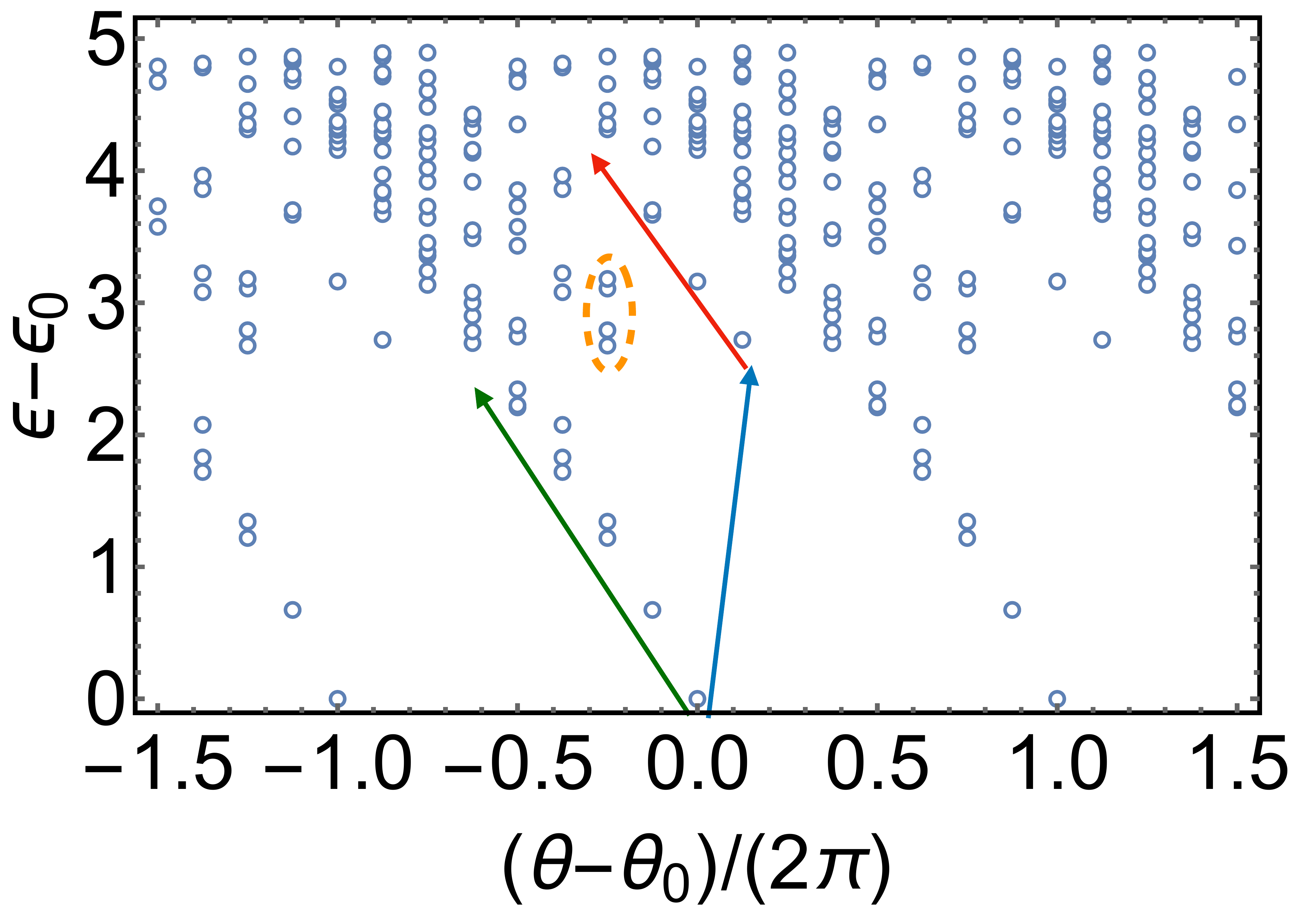}
    \caption{Relative energies, $\epsilon_n - \epsilon_0$, and crystal momenta, $\theta_n - \theta_0$, for the lowest 100 energy eigenstates  of $H_{\text{nc}}$ (\ref{pert}) for $(k_1, k_2) = (1,5)$ and $N = 8$. The spectrum resembles the chiral boson theory $H_0'$ (\ref{eq:free}): the states along the guiding green line can be identified with the left-moving mode $\phi_2$, while the states along the blue line correspond to the right-moving mode $\phi_1$ and the states along the red line correspond to composites of left-moving excitations and one right-moving excitation. The circled states can be identified with the $4$ degenerate zero mode excitations of $H_0'$ with charges $Q_1, Q_2 = \pm 1$. For clarity, results are plotted in a repeated zone scheme where $\theta_n - \theta_0$ ranges from $-3\pi$ to $3\pi$. 
}
  \la{fig:spectrum15}
\end{figure}

Taking advantage of this additional quantum number, we compute the energy difference $\epsilon_n -\epsilon_0$ and crystal momentum difference $\theta_n - \theta_0$ for the lowest 100 energy eigenstates $\{|\Psi_n\>\}$ that belong to the same fermion parity sector as the ground state $|\Psi_0\>$. (Here $\epsilon_0, \theta_0$ denote the energy and crystal momentum of the ground state). We perform this computation for $N = 8$, which corresponds to $16$ impurities. To find the crystal momentum difference $\theta_n - \theta_0$, we use the following identity:
\begin{align}
e^{i (\theta_{m} - \theta_n)}= \frac{\langle \Psi_m |U_3|\Psi_n\rangle}{\langle \Psi_m |U_1|\Psi_n\rangle} .
\label{crymomid}
\end{align}


We first present our results in the case $(k_1, k_2) = (1,5)$, since they are easier to interpret. In this case, the low energy spectrum of $H_{\text{nc}}$ closely resembles that of a standard chiral boson edge theory with two counter-propagating modes, namely
\begin{align}
	H_{0}' = \frac{1}{4\pi}\int_{0}^{L}[v_1k_1(\partial_{x}\phi_{1}(x))^{2}+v_2k_2(\partial_{x}\phi_{2}(x))^{2}]dx
	\label{eq:free}
\end{align}
where $\phi_1, \phi_2$ are chiral boson fields obeying the commutation relations (\ref{eq:comm}), and where $(k_1, k_2) = (1, 5)$. (Note that $H_{0}'$ is identical to the clean edge theory $H_0$ (\ref{eq:hzero}) except that it has different velocities $v_1 \neq v_2$ for the two counterpropagating modes). 

To see the similarity between the low energy spectrum of $H_{\text{nc}}$ and $H_{0}'$, consider the green, blue, and red lines in Fig.~\ref{fig:spectrum15}. Along the green line, we see a collection of states with a linear dispersion with a negative slope. We also see approximate degeneracies of $1,1,2,3,5...$. This dispersion and degeneracy counting exactly matches the left-moving phonon modes associated with $\phi_2$. Likewise, along the blue line in Fig.~\ref{fig:spectrum15}, we can see two non-degenerate states with a much steeper positive slope. This dispersion matches the right-moving phonon modes associated with $\phi_1$ if we take the $v_1$ to be larger than $v_2$, or more specifically $v_1/v_2 \approx 5$. Likewise, along the red line in Fig.~\ref{fig:spectrum15}, we see another set of states with a linear dispersion and a negative slope, and with approximate degeneracies of $1, 1, 2,...$. This dispersion and degeneracy counting is consistent with the collection of phonon excitations that are made up of the lowest energy excitation of $\phi_1$ together with multiple phonon excitations of $\phi_2$. Finally, note the cluster of $4$ approximately degenerate states circled in Fig.~\ref{fig:spectrum15}. These states match the lowest energy \emph{zero-mode} excitations of $H_0'$ which carry charges $Q_1 = \pm 1$ and $Q_2 = \pm 1$, and come in a $4$-fold multiplet. 
\begin{figure}[t]
  \centering
\includegraphics[width=0.9\columnwidth]{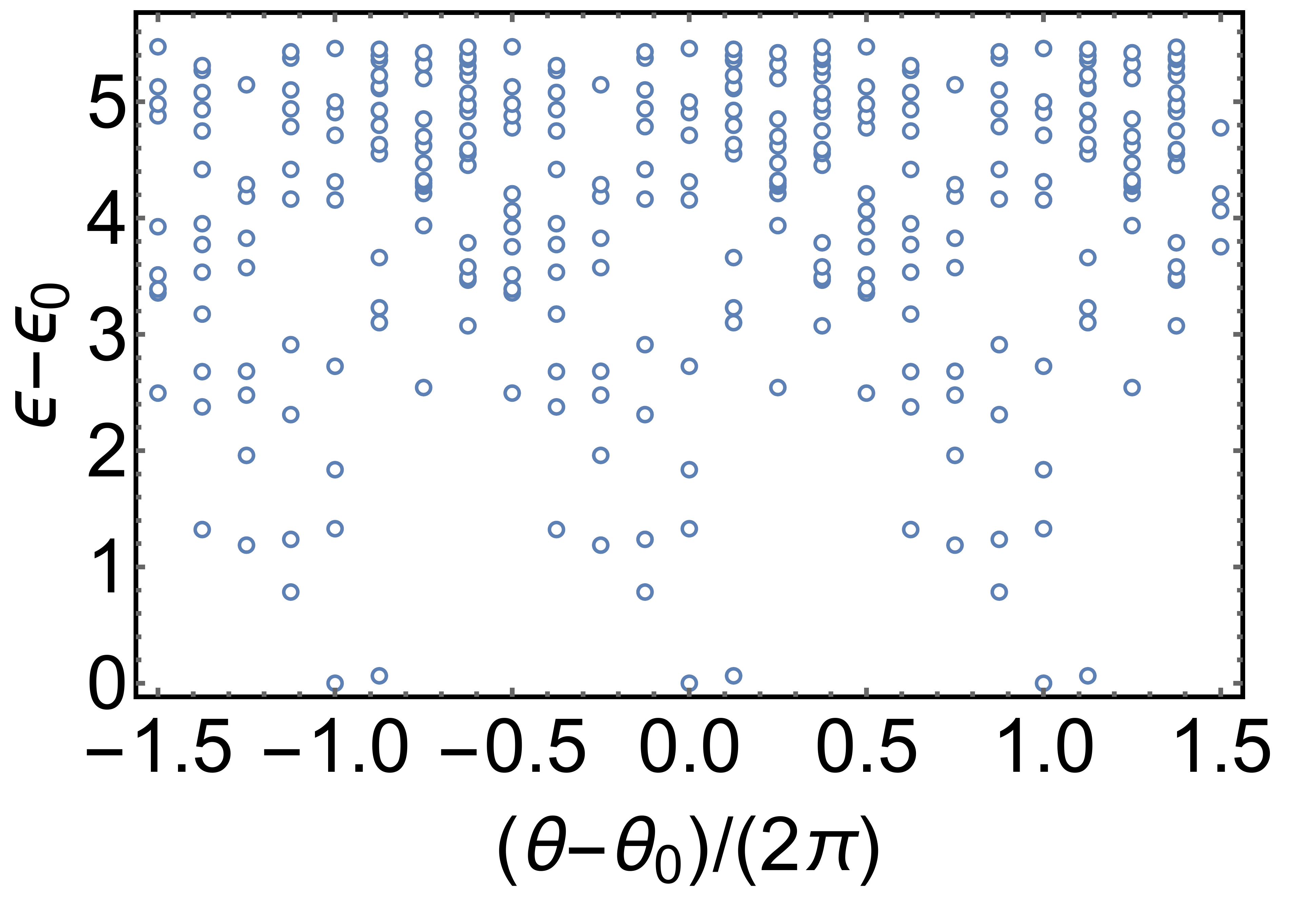}
    \caption{Relative energies, $\epsilon_n - \epsilon_0$, and crystal momenta, $\theta_n - \theta_0$, for the lowest 100 energy eigenstates of $H_{\text{nc}}$ for $(k_1, k_2) = (1,3)$ and $N = 8$.  For clarity, results are plotted in a repeated zone scheme where $\theta_n - \theta_0$ ranges from $-3\pi$ to $3\pi$. }
  \la{fig:spectrum13}
\end{figure}

Combining all of this numerical evidence, including the scaling of the gap $\Delta$, it seems likely that the low energy physics of $H_{\text{nc}}$ is indeed described by the chiral boson edge theory $H_0'$. We have thus come full circle: we started with a chiral boson edge theory $H_0$; we then added strong impurity scattering, which resulted in an energy gap and an extensive ground state degeneracy. Finally, we added a further perturbation $H_{\text{nc}}$, splitting the ground state degeneracy and leading to $H_0'$, which is essentially the same chiral boson edge theory that we started with, but at a lower energy scale. 

We now move on to the case $(k_1, k_2) = (1,3)$, shown in Fig. \ref{fig:spectrum13}. In this case, we have not been able to identify any structure associated with the low energy spectrum of $H_{\text{nc}}$, and we have not found a candidate field theory that matches it. The only conclusion we can draw, coming primarily from Fig. \ref{fig:gapscaling}, is that the gap appears to vanish in the thermodynamic limit. A more systematic numerical study may be necessary to understand this example.

\section{Discussion}
\label{sec:disc}

In this paper, we have constructed a family of finite dimensional edge theories describing the boundaries of Abelian topological phases with $K$-matrices of the form $K = \bpm k_1 & 0 \\ 0 & -k_2 \epm$. Importantly this family includes both type-I and type-II topological phases, i.e.~phases with both gappable and ungappable boundaries and with vanishing thermal Hall coefficient, $\kappa_H = 0$. These edge theories are defined by a collection of algebraic relations (\ref{alg0}-\ref{alg4}) satisfied by the elementary local operators $U_i$. 

An interesting aspect of our edge theories (for $k_1 \neq k_2$) is that they do not seem to have a tensor product structure. More precisely, our edge theories do not have an obvious description as a tensor product Hilbert space with constraints (e.g.~like the toric code edge theory mentioned in the introduction). This lack of a tensor product structure is particularly intriguing in the case of the type-II topological phases where $k_1 k_2$ is not a perfect square. We are not aware of \emph{any} edge theory for these phases with a (constrained) tensor product structure, so one can reasonably conjecture that this is a general feature of type-II topological phases. This conjecture is reminiscent of an observation of Jones and Metlitski that some symmetry protected topological phases do not support edge theories with tensor product Hilbert spaces~\cite{jones20191d}. Here, the conjecture about type-II topological phases is even stronger, both because it doesn't depend on symmetry, and because it rules out tensor product Hilbert spaces with \emph{constraints}.

While we have focused on a particular class of $2 \times 2$ fermionic $K$-matrices, our construction can be readily extended to any $2m \times 2m$ bosonic or fermionic $K$-matrix with vanishing thermal Hall coefficient, $\kappa_H = 0$. In the general case, one would start with the standard chiral boson edge theory~\cite{wen1995topological} with $2m$ fields $\phi_1,...,\phi_{2m}$ and then introduce $2m$ different impurity scattering terms of the form $ U\cos(\Lambda^T K \Phi)$ where $\Lambda$ is a $2m$ component integer vector and $\Phi = (\phi_1,..., \phi_{2m})^T$. For an appropriate choice of scattering terms, the resulting impurity model should have a phonon gap, and the ground state subspace of this model can then be used to derive a finite dimensional edge theory, as we did here. In this way, one can construct a finite dimensional edge theory for any Abelian topological phase with $\kappa_H = 0$. On the other hand, our construction does not have an obvious generalization to \emph{non-Abelian} topological phases.

We see a number of directions for future work. One direction would be to prove rigorously that our type-II edge theories cannot be gapped by any local Hamiltonian. Since our edge theories are finite dimensional, they provide a particularly convenient setting for making precise statements of this kind. 

Another direction would be to use our edge theories as a platform for numerical investigations of type-II edges. Such numerical studies could help us understand what kinds of gapless energy spectra are possible for type-II edges. For example, are there general bounds on the energy gap or density of states as a function of system size?

It would also be interesting to use our edge theories to build lattice models for type-II topological phases. For example, the coupled wire construction of Ref.~\onlinecite{kane2002coupled} provides a general method for constructing models for bulk topological phases starting from edge theories. In our case, since our edge theories are finite dimensional, it may be possible to construct a fully microscopic lattice model using such an approach.

\acknowledgments

This work was supported in part by the Simons Collaboration on Ultra-Quantum Matter, which is a grant from the Simons Foundation (651440, ML). SG was supported by NSF CAREER Grant No. DMR-1944967.

\appendix

\section{Derivation of Eqs.~(\ref{alg0}-\ref{alg4})}
\label{derivrelapp}
In this Appendix, we derive the basic algebraic relations obeyed by the $U_i$ operators, namely Eqs.~(\ref{alg0}-\ref{alg4}). We reprint these equations below for convenience:
\begin{align}
&U_i^{-1} = U_i^\dagger \label{alg0app} \\
&U_i U_j =e^{2\pi i \mathcal{Z}^{-1}_{ij}} U_j U_i \label{alg1app} \\
&U_i^{\beta} U_{i+1}^{-2\alpha}U_{i+2}^{\beta}=\mathbb{I} \label{alg2app} \\
&U_1 U_2 \cdots U_{2N} = e^{i \theta_1} \mathbb{I}, \label{alg3aapp} \\
& U_1^{-1} U_2 U_3^{-1} \cdots U_{2N} =e^{i\theta_2} \mathbb{I} \label{alg3bapp} \\
&\text{Tr}\left(U_2 U_4 \cdots U_{2N}\right) = 0 \label{alg4app} 
\end{align}
Our derivation is based on an analogous set of relations for the $\Gamma_i$ operators:
\begin{align}
[\Gamma_i, \Gamma_j] &= -2\pi i \mathcal{Z}^{-1}_{ij} \label{Gammacomm} \\
\beta \Gamma_i - 2\alpha \Gamma_{i+1} + \beta \Gamma_{i+2} &= C_{i+2}-C_{i} \label{Gammaab} \\
\sum_{i=1}^{2N} \Gamma_i &= -C_{2N+1} \label{Gammaplus} \\
\sum_{i=1}^{2N} (-1)^i \Gamma_i &= -C_{2N+2} \label{Gammaalt}
\end{align}
Here, Eq.~(\ref{Gammacomm}) follows immediately from $[C_i, C_j] = 2\pi i \mathcal{Z}_{ij}$ together with $\Gamma_i = \sum_j \mathcal{Z}^{-1}_{ji} C_j$. Likewise,
(\ref{Gammaab}-\ref{Gammaalt}) follow from $C_i = \sum_j \mathcal{Z}_{ji} \Gamma_j$, along with the explicit form of the $\mathcal{Z}_{ij}$ matrix (\ref{zij}). We note that (\ref{Gammaab}) only holds for $i = 1,...,2N-2$; for the special cases where $i = 2N-1$ or $2N$, the relations take the modified form
\begin{align}
\beta \Gamma_{2N-1} &- 2\alpha \Gamma_{2N} + \beta \Gamma_{1} = \nonumber \\
&C_{1}-C_{2N-1}+(\alpha-\beta)C_{2N+1} + (\alpha+\beta)C_{2N+2}, \nonumber \\
\beta \Gamma_{2N} &- 2\alpha \Gamma_{1} + \beta \Gamma_{2} = \nonumber  \\
&C_{2}-C_{2N}+(\alpha-\beta)C_{2N+1} -(\alpha+\beta)C_{2N+2}, 
\label{Gammaab2}
\end{align}

We now use (\ref{Gammacomm}-\ref{Gammaab2}), together with the definition $U_i = e^{i \Gamma_i + i \phi}$, to derive Eqs.~(\ref{alg0app}-\ref{alg4app}). We start with Eq.~(\ref{alg1app}), the first nontrivial relation. This relation follows immediately from (\ref{Gammacomm}) together with the Baker-Campbell-Hausdorff formula.  To derive Eq.~(\ref{alg2app}), we use (\ref{Gammaab}), (\ref{Gammaab2}) together with the fact that $e^{i C_i} = \mathbb{I}$ within the ground state subspace to deduce that 
\begin{align}
e^{i(\beta \Gamma_i - 2\alpha \Gamma_{i+1} + \beta \Gamma_{i+2})}= e^{i \beta \pi} \mathbb{I}
\label{alg3gamma}
\end{align}
Here, the phase factor on the right hand side comes from the Baker-Campbell-Hausdorff formula. Decomposing the left hand side of (\ref{alg3gamma}) into a product of exponentials, we obtain
\begin{align}
e^{i\beta \Gamma_i} e^{- i 2\alpha \Gamma_{i+1}} e^{i \beta \Gamma_{i+2}} = e^{i\varphi} \mathbb{I}
\label{varphiid}
\end{align}
where 
\begin{align*}
\varphi = \frac{\pi}{4} (k_2 - k_1)^2 \mathcal{Z}^{-1}_2 - \pi(k_2^2 - k_1^2) \mathcal{Z}^{-1}_1 + \frac{\pi (k_2 - k_1)}{2}
\end{align*} 
Here, we are using the abbreviation $\mathcal{Z}^{-1}_{j} \equiv \mathcal{Z}^{-1}_{i (i+j)}$. Eq.~(\ref{alg2app}) then follows from (\ref{varphiid}), provided that we choose the phase $\phi$ in the definition of $U_i$ to be $\phi = \frac{\varphi}{2k_1}$, that is:
\begin{align}
\phi = \frac{\pi}{8k_1} (k_2 - k_1)^2 \mathcal{Z}^{-1}_2 - \frac{\pi(k_2^2 - k_1^2)}{2 k_1} \mathcal{Z}^{-1}_1 + \frac{\pi (k_2 - k_1)}{4k_1}
\label{phidef}
\end{align}

Next we derive Eqs.~(\ref{alg3aapp}-\ref{alg3bapp}). This derivation follows the same logic as Eq.~(\ref{alg2app}). First, we use (\ref{Gammaplus})-(\ref{Gammaalt}) together with the fact that $e^{iC_i} = \mathbb{I}$ within the ground state subspace, to deduce that
\begin{align}
\exp \left(i \sum_{i=1}^{2N} \Gamma_i \right) = \exp \left(i \sum_{i=1}^{2N} (-1)^i \Gamma_i \right) = \mathbb{I}
\label{globalgamma}
\end{align}
Decomposing the left hand side into a product of exponentials using the Baker-Campbell-Hausdorff formula, it is straightforward to show that (\ref{alg3aapp}-\ref{alg3bapp}) holds with
\begin{align}
\theta_1 &= \pi \sum_{j=1}^{2N-1} (2N-j) \mathcal{Z}^{-1}_{j} + 2N \phi  \nonumber \\
\theta_2 &= \pi \sum_{j=1}^{2N-1} (2N-j) (-1)^{j} \mathcal{Z}^{-1}_{j}
\label{thetadef}
\end{align}

All that remains is Eq.~(\ref{alg4app}). To derive this relation, it suffices to show that 
\begin{align}
\text{Tr}_\mathcal{H}\left(e^{i \sum_{i=1}^{2N} \Gamma_{2i}}\right) = 0
\end{align}
or equivalently
\begin{align}
\text{Tr}_\mathcal{H}\left(e^{-i (C_{2N+1} + C_{2N+2})/2}\right) = 0
\label{trcid}
\end{align}
since $\sum_{i=1}^{2N} \Gamma_{2i} = - (C_{2N+1} + C_{2N+2})/2$ by (\ref{Gammaplus}-\ref{Gammaalt}). To show (\ref{trcid}), consider the ``electron'' operator $\psi_{1/2}$ defined in (\ref{electrondef1}) below.
One can check that $\psi_{1/2}$ commutes with $e^{i C_j}$ for all $j$ and therefore preserves the ground state subspace $\mathcal{H}$. At the same time, $\psi_{1/2}$ \emph{anticommutes} with $e^{-i (C_{2N+1} + C_{2N+2})/2}$. Eq.~(\ref{trcid}) follows immediately from these properties (see e.g. Lemma~\ref{tracelemma} in App.~\ref{sec:deg}).

\section{Electron operators}
\label{electronapp}
In this Appendix, we construct a collection of fermion parity-odd operators within our edge theory. These operators are local (in the fermionic sense) so we will refer to them as ``electron'' operators.

\subsection{Microscopic definition of electron operators}
We begin with a microscopic definition of the electron operators. We denote these operators by $\{\psi_{i+1/2}\}$ where $i$ runs from $0,...,2N-1$. Note that we index the electron operators by \emph{half}-integers instead of integers; this turns out to be convenient notation as it leads to more symmetrical algebraic relations (\ref{psisq}-\ref{psipsicomm}). We will think of $\psi_{i+1/2}$ as living halfway between the $i$th and $(i+1)$st impurity. 

We start by defining the electron operator $\psi_{1/2}$, which lives between impurities $2N$ and $1$. We define\footnote{Alternatively, we could define our electron operators using $\Gamma_{2N+2}$ instead of $\Gamma_{2N+1}$; this would give rise to a slightly different set of fermion parity-odd operators.}
\begin{align}
\psi_{1/2} = e^{i \Gamma_{2N+1} + i \phi'}
\label{electrondef1}
\end{align}
where $\Gamma_{2N+1}$ is defined like $\Gamma_1,...,\Gamma_{2N}$, i.e.
\begin{align}
\Gamma_{2N+1} = \sum_{j=1}^{2N+2} \mathcal{Z}_{j(2N+1)}^{-1} C_j
\end{align}
Here $\phi'$ is an additional phase which we include to simplify the algebraic relations obeyed by the electron operators [in particular (\ref{psisq})]. Specifically, we choose
\begin{align}
\phi' = -\frac{\pi}{4}(k_1+k_2) - \frac{\pi}{8} (k_2-k_1)^2 \mathcal{Z}^{-1}_1
\end{align}
Like the $U_i$ operators, it is easy to check that $\psi_{1/2}$ commutes with $a_{m,\theta}$ and $\cos(C_j)$ and therefore defines a legitimate operator acting within the ground state subspace $\mathcal{H}$. It is also easy to see that $\psi_{1/2}$ anticommutes with the fermion parity operator $e^{i(C_{2N+1} + C_{2N+2})/2}$ and therefore carries odd fermion parity. 
 
To define the other electron operators $\psi_{n+1/2}$, with $n=1,...,2N-1$, we multiply $\psi_{1/2}$ by an appropriate string operator:
\begin{align}
\psi_{n+1/2} = e^{i (n \nu +\phi') } \psi_{1/2} \cdot (U_1^{-k_1} U_2^{-k_1} \cdots U_n^{-k_1}) 
\label{electrondef2}
\end{align}
Here $\nu$ is the following phase factor, which we again include to simplify the relations below [namely (\ref{psisq})]:
\begin{align}
\nu =  \pi k_1 \mathcal{Z}^{-1}_{1(2N+1)} &+ \frac{\pi}{4} \mathcal{Z}^{-1}_2 (k_2 - k_1)^2 \nonumber \\
&- \frac{\pi}{4} \mathcal{Z}^{-1}_1 (3k_2+k_1)(k_2 -k_1) 
\end{align}
One might worry that $\psi_{n+1/2}$ is a nonlocal operator given its appearance. However, we will see below that $\psi_{n+1/2}$ is in fact local (in the fermionic sense). The string operator that appears in the definition of $\psi_{n+1/2}$ should be thought of as analogous to a Jordan-Wigner string.

Relatedly, notice that $U_1^{-k_1} U_2^{-k_1} \cdots U_n^{-k_1}$ is equivalent (up to a phase) to the anyonic string operator $(W_{ab}^1)^{-k_1}$ defined in Sec.~\ref{sec:anyonstring}. This makes sense: the operator $W_{ab}^1$ moves an anyon from $a$ to $b$, and a composite of $k_1$ such anyons is a local fermion (i.e. electron). Thus, $(W_{ab}^1)^{-k_1}$ is a string operator that moves an electron from $a$ to $b$.

\subsection{Algebraic definition of electron operators}
An alternative way to define the electron operators $\psi_{n+1/2}$ is by their algebraic relations. The defining algebraic relations for the electron operators are as follows: 
\begin{align}
\psi_{i+1/2}^2 &= U_i^\beta U_{i+1}^{-\beta} \label{psisq}\\
\psi_{i-1/2}^{-1} \psi_{i+1/2} &= e^{i \nu} U_i^{-k_1} , \quad i=1,...,2N-1\label{psipsi}\\
U_i \psi_{j+1/2} &= e^{2\pi i \mathcal{Z}^{-1}_{(i-j)(2N+1)}} \psi_{j+1/2} U_i \label{psiUcomm} \\
\psi_{i+1/2} \psi_{j+1/2} &= -e^{-\pi i k_1 \beta \mathcal{Z}^{-1}_{ij}} \psi_{j+1/2} \psi_{i+1/2}, \ \ i \neq j \label{psipsicomm}
\end{align}
where the difference $i-j$ is defined modulo $2N$. Here, Eq.~(\ref{psisq}) follows from a Baker-Campbell-Hausdorff calculation similar to those discussed in Appendix~\ref{derivrelapp}, while Eq.~(\ref{psipsi}) follows immediately from the definition (\ref{electrondef2}). The last two equations (\ref{psiUcomm}-\ref{psipsicomm}) also follow from Baker-Campbell-Hausdorff calculations, with the help of the identities (\ref{z2nplus1form1}) and (\ref{eq:zij0}). 

To fully understand (\ref{psisq}-\ref{psipsicomm}), it is useful to have a more explicit formula for the matrix elements $\mathcal{Z}_{j(2N+1)}^{-1}$ with $1 \leq j \leq 2N$. A straightforward linear algebra calculation gives
\begin{align}
\mathcal{Z}_{j(2N+1)}^{-1} = k_1 (\mathcal{X}^{-1}_{j1} + \mathcal{X}^{-1}_{j(2N)}) 
\label{z2nplus1form1}
\end{align}
where $\mathcal{X}$ is the $2N \times 2N$ matrix defined in Eq.~(\ref{Xdef}). Substituting the formula for $\mathcal{X}^{-1}$ (\ref{xinv}), we obtain the following expression for $\mathcal{Z}_{j(2N+1)}^{-1}$:
\begin{align}
\mathcal{Z}_{j(2N+1)}^{-1} = \frac{2 k_1}{k_2 -k_1} \frac{x^{N+1-j} + x^{j-1-N} +x^{N-j} + x^{j-N} }{(x^{N} - x^{-N})(x-x^{-1})}
\label{z2nplus1form2}
\end{align}

With these formulas, we are now ready to discuss some important features of (\ref{psisq}-\ref{psipsicomm}). Our first comment is about the commutation relation (\ref{psiUcomm}): to understand the structure of this commutation relation note that $\mathcal{Z}_{j(2N+1)}^{-1}$ is exponentially small for large $|j|$, $|2N-j|$ (since $\mathcal{X}^{-1}$ is a quasidiagonal matrix). It follows that $U_i$ and $\psi_{j+1/2}$ approximately commute at large separations. This is important because it means that $\psi_{j+1/2}$ are indeed local operators (in the fermionic sense), as we claimed earlier. 

Our second comment is about the last relation (\ref{psipsicomm}): note that this equation implies that $\psi_{i+1/2}$ and $\psi_{j+1/2}$ approximately \emph{anti-commute} at large separations since $\mathcal{Z}^{-1}_{ij}$ is exponentially small for large $|i-j|$. This makes sense since the $\psi_{i+1/2}$ describe fermionic operators.

Finally, it is worth mentioning that the relations (\ref{psisq}-\ref{psipsicomm}) simplify substantially in the special case $k_1 = k_2 = k$. In this case (\ref{psisq}-\ref{psipsicomm}) reduce to
\begin{align}
\psi_{i+1/2}^2 &= \mathbb{I} \\
\psi_{i-1/2} \psi_{i+1/2} &= e^{i k \pi/2} U_i^{-k} , \quad i=1,...,2N-1 \\
U_i \psi_{j+1/2} &= - \psi_{j+1/2} U_i, \quad i = j, j+1 \\
\psi_{i+1/2} \psi_{j+1/2} &= -\psi_{j+1/2} \psi_{i+1/2}, \quad i \neq j 
\end{align}

 \section{Matrix representation of $U_i$}
 \label{sec:numerical}
In this Appendix, we describe how to construct an explicit matrix representation of the $U_i$ operators using the formalism of Ref.~\onlinecite{ganeshan2016formalism}. 

The simplest way to construct a representation is to use our original expression for $U_i$ in terms of $C_j$:
\begin{align}
U_i = \exp\left(i \sum_{j=1}^{2N+2} \mathcal{Z}^{-1}_{ji} C_j + i \phi \right)
\label{UiC}
\end{align}
Here $\phi$ is given in Eq.~(\ref{phidef}), and the $C_i$'s obey
\begin{align}
[C_i, C_j] = 2\pi i \mathcal{Z}_{ij}, \quad e^{i C_i} = \mathbb{I}
\label{Cirel}
\end{align}
In principle, Eqs.~(\ref{UiC}-\ref{Cirel}) completely determine the form of the $U_i$ operators, but they are not easy to work with since $\mathcal{Z}_{ij}$ is a complicated skew-symmetric matrix. We now make a change of variables to simplify these relations. Specifically, we define
\begin{align}
C_i' = \sum_{j=1}^{2N+2} \mathcal{V}_{ij} C_j + \chi_i
\end{align}
for some matrix $\mathcal{V}$ and some vector $\chi$ that we will choose below. Then,  $[C_i', C_j'] = 2\pi i \mathcal{Z}'_{ij}$ where
\begin{align}
\mathcal{Z}' = \mathcal{V} \mathcal{Z} \mathcal{V}^T
\end{align}
We choose $\mathcal{V}$ to be a matrix with integer entries and determinant $\pm 1$, with the property that $\mathcal{Z}'$ takes the simple form
\begin{align}
\mathcal{Z}' = \bpm 0_{N+1} & - \mathcal{D} \\ \mathcal{D} & 0_{N+1} \epm, \ \ \mathcal{D} = \bpm d_1 & 0 & \cdots & 0 \\
0 & d_2 & \cdots & 0 \\ \vdots & \vdots & \vdots & \vdots \\ 0 & 0 & \cdots & d_{N+1} \epm
\label{skewnorm}
\end{align}
where $d_1,...,d_{N+1}$ are positive integers. Here, the matrix $\mathcal{V}$ is an integer change of basis that puts $\mathcal{Z}$ into ``skew-normal'' form. Such a change of basis always exists, but it is not unique~\cite{NewmanBook}.
 
After finding $\mathcal{V}$, we then choose the offset $\chi$ so that
\begin{align}
\chi_i = \pi \sum_{j < k} \mathcal{V}_{ij} \mathcal{V}_{ik} \mathcal{Z}_{jk} \pmod{2\pi}
\end{align}
This choice ensures that $e^{i C_i'} =\mathbb{I}$, as one can verify using the Baker-Campbell-Hausdorff formula, together with $e^{i C_j} = \mathbb{I}$.

Next we define $\widetilde{C}'_i$ by rescaling $C'_i$:
\begin{align}
C'_i = \sum_{j=1}^{2N+2} \widetilde{\mathcal{D}}_{ij} \widetilde{C}'_j, \quad
\widetilde{\mathcal{D}}_{ij} = \bpm \mathcal{D} & 0_{N+1} \\ 0_{N+1} & \mathcal{D} \epm
\end{align}
Putting this together, we can write $U_i$ in terms of $\widetilde{C}_j'$ as
\begin{align}
U_i = \exp\left(i \sum_{j=1}^{2N+2} Y_{ij} \widetilde{C}'_j + i\tilde{\chi}_i + i \phi\right)
\end{align}
where the matrix $Y_{ij}$ and the vector $\tilde{\chi}_i$ are defined by
\begin{align}
Y =  -\mathcal{Z}^{-1} \mathcal{V}^{-1} \widetilde{\mathcal{D}}, \quad \tilde{\chi} = \mathcal{Z}^{-1} \mathcal{V}^{-1} \chi
\end{align}
By construction, the $\widetilde{C}_i'$ operators obey  
\begin{align}
[\widetilde{C}'_i, \widetilde{C}'_{i+N+1}] = -\frac{2\pi i}{d_i},
\label{ciprrel1}
\end{align}
with all other commutators vanishing. Also
\begin{align}
e^{i d_i \widetilde{C}_i'} = e^{i d_i \widetilde{C}'_{i+N+1}} = \mathbb{I} 
\label{ciprrel2}
\end{align}
This completes our change of variables from $C_i$ to $\widetilde{C}'_i$. 

The advantage of the new variables is that it is easy to find a representation for $e^{i\widetilde{C}'_i}$. Indeed, given the algebra (\ref{ciprrel1}-\ref{ciprrel2}), it is clear that each pair of operators $e^{i \widetilde{C}'_i}$ and $e^{i \widetilde{C}'_{i+N+1}}$ can be represented as $d_i \times d_i$ clock and shift matrices: 
\begin{align}
e^{i \widetilde{C}'_i} = A_{d_i}, \quad e^{i \widetilde{C}'_{i+N+1}} = B_{d_i}
\label{ciprrep}
\end{align}
where $A_d$ and $B_d$ are defined by
\begin{align}
A_{d} &= \bpm 1 & 0 & \cdots & 0 \\ 
0 & e^{\frac{i 2 \pi}{d}} & \cdots & 0 \\ 
\vdots & \vdots & \vdots  & \vdots \\
 0 & 0 & \cdots & e^{\frac{i 2\pi (d-1)}{d}} \epm, \nonumber \\
B_{d} &= \bpm 0 & 0 & \cdots & 0 & 1 \\ 
1 & 0 & \cdots & 0 & 0\\ 
0 & 1 & \cdots & 0 & 0\\
\vdots & \vdots & \vdots & \vdots & \vdots \\
 0 & 0 & \cdots & 1 & 0 \epm
\end{align} 

To translate this into a representation for $U_i$, note that the matrix $Y_{ij}$ has integer entries, as one can easily verify using the fact that $\mathcal{V} \mathcal{Z} \mathcal{V}^T$ is of the form given in (\ref{skewnorm}). Therefore each $U_i$ is a product of integer powers of $e^{i \widetilde{C}_j'}$. Specifically, we can write $U_i$ as
\begin{align}
U_i = e^{i (\tilde{\chi_i} + \nu_i + \phi)} \prod_{j=1}^{N+1} \left(e^{i Y_{ij} \widetilde{C}'_j} e^{i Y_{i (j+N+1)} \widetilde{C}'_{j+N+1}}\right)
\end{align}
where the extra phase $\nu_i$ comes from the Baker-Campbell-Hausdorff formula, and is given by
\begin{align}
\nu_i = -\pi \sum_{j=1}^{N+1} \frac{Y_{ij} Y_{i (j+N+1)}}{d_i}
\end{align} 

Plugging in (\ref{ciprrep}), we obtain the following represention for $U_i$ as a tensor product of clock and shift matrices:
\begin{align}
 U_i = e^{i (\tilde{\chi_i} + \nu_i + \phi)} \bigotimes_{j=1}^{N+1} \left(A_{d_j}^{Y_{ij}} B_{d_j}^{Y_{i (j+N+1)}}\right)
\label{Uiclockform}
\end{align}
This is the desired representation of $U_i$.

To illustrate this construction, consider the case where $(k_1, k_2) = (1,3)$. In this case, one finds that the $d_i$'s defined by (\ref{skewnorm}) are 
\begin{align}
d_1 = d_2 = ... = d_N = 1, \quad \quad d_{N+1}  = D_N
\end{align}
This means that all the clock and shift matrices in the tensor product (\ref{Uiclockform}) are trivial (i.e.~equal to $1$) except for those labeled by $j = N+1$. Hence (\ref{Uiclockform}) reduces to
\begin{align}
U_i = e^{i (\tilde{\chi_i} + \nu_i + \phi)} A_{D_N}^{a_i} B_{D_N}^{b_i} 
\end{align}
where $a_i = Y_{i(N+1)}$ and $b_i = Y_{i (2N+2)}$ are integers, and $A_{D_N}, B_{D_N}$ are $D_N \times D_N$ clock and shift matrices. 

Another illustrative example is $(k_1, k_2) = (1, 9)$. In this case one finds
\begin{align}
d_1 = d_2 = ... = d_N = 2, \quad \quad d_{N+1} = Q
\end{align}
where $Q = \frac{D_N}{2^N} = \frac{1}{3}(2^N-1)$. Hence $U_i$ is given by a tensor product of $N+1$ matrices, of which $N$ are $2 \times 2$ Pauli operators, and one of which is a $Q \times Q$ matrix built out of clock and shift matrices raised to integer powers. Note that in both of these examples, the Hilbert space does not have a \emph{local} tensor product structure (e.g. like a spin chain) since the $(N+1)$st block has an exponentially large dimension $d_{N+1}$.

\section{Algebra of anyonic string operators}
\label{sec:nonlocal}
 In this Appendix, we derive the commutation algebra of the two string operators
	 \begin{align*}
		W^1_{ab}=\prod_{i=a}^b U_i,\quad 
		W^2_{ab}=\prod_{i=a}^b U_i^{(-1)^{i}},
	\end{align*}
Specifically, we show that for an ``interleaved'' geometry with $a < c < b < d$, the string operators obey the commutation relations
\begin{align}
	W^i_{ab}W^j_{cd}=e^{i \alpha_{ij}}W^j_{cd} W^i_{ab},
\end{align}
where
\begin{align}
\alpha_{11} \approx \frac{2\pi}{k_1}, \quad \quad \alpha_{22} \approx -\frac{2\pi}{k_2}, \quad \quad
\alpha_{12} \approx \alpha_{21} \approx 0 
\end{align}
up to errors of order $\mathcal{O}(x^{-\min(|a-c|, |b-c|, |a-d|, |b-d|)})$.

We start by evaluating $\alpha_{11}$. Using (\ref{alg1}), we have
\begin{align}
		\alpha_{11}=2\pi \sum_{i=a}^b\sum_{j=c}^d \mathcal{Z}^{-1}_{ij}
\end{align}
Substituting the formula (\ref{Zapprox}) for $\mathcal{Z}^{-1}_{ij}$ in the limit $N\rightarrow \infty$, we obtain
\begin{align}
\alpha_{11}= \frac{4\pi}{k_2-k_1}\sum_{i=a}^b\sum_{j=c}^d x^{-|i-j|}\sgn(j-i)
\label{a11form}
\end{align}		
Evaluating the double sum gives
\begin{align}
\sum_{i=a}^b\sum_{j=c}^d x^{-|i-j|}&\sgn(j-i)  \nonumber \\
& = \frac{(x-x^{a-c+1})-(x^{c-d}-x^{a-d})}{(x-1)^2} \nonumber \\
&+ \frac{(x-x^{c-b})-(x^{b-d+1}-x^{c-d})}{(x-1)^2}
	\end{align}
Neglecting terms of order $\mathcal{O}(x^{-\min(|a-c|, |b-c|, |a-d|, |b-d|)})$ gives
\begin{align}
\sum_{i=a}^b\sum_{j=c}^d x^{-|i-j|}\sgn(j-i) \approx \frac{2x}{(x-1)^2}
\end{align}
so that
\begin{align}
\alpha_{11} &\approx  \frac{8 \pi x}{(k_2-k_1)(x-1)^2} \nonumber \\
&= \frac{8 \pi}{(k_2-k_1)(x+x^{-1}-2)} \nonumber \\
&= \frac{2\pi}{k_1}
\label{a11form2}
\end{align}
where the third equality follows from 
\begin{align}
x+x^{-1} = \frac{2(k_1+k_2)}{k_2-k_1}
\label{xinvx}
\end{align}

We can evaluate $\alpha_{22}$ in a similar fashion. First we note that
\begin{align}
\alpha_{22}&=2\pi \sum_{i=a}^b\sum_{j=c}^d (-1)^{i+j}\mathcal{Z}^{-1}_{ij} \nonumber \\
&= \frac{4\pi}{k_2-k_1}\sum_{i=a}^b\sum_{j=c}^d (-x)^{-|i-j|}\sgn(j-i)
\end{align}
We then note that the latter formula is identical to the one for $\alpha_{11}$ (\ref{a11form}) except with $x$ replaced with $-x$. Making this replacement in (\ref{a11form2}), we deduce that 
\begin{align}
\alpha_{22} \approx   \frac{8\pi}{(k_2 -k_1)(-x-x^{-1}-2)} = -\frac{2\pi}{k_2}
\end{align}
where we again use the identity (\ref{xinvx}).

Next, consider $\alpha_{21}$. We have
\begin{align}
\alpha_{12}&=2\pi \sum_{i=a}^b\sum_{j=c}^d (-1)^{i}\mathcal{Z}^{-1}_{ij} \nonumber \\
&= \frac{4\pi}{k_2-k_1}\sum_{i=a}^b\sum_{j=c}^d (-1)^i x^{-|i-j|}\sgn(j-i)
\end{align}
Evaluating the double sum gives
\begin{align}
&\sum_{i=a}^b\sum_{j=c}^d (-1)^i x^{-|i-j|}\sgn(j-i)  \nonumber \\
& = \frac{(-1)^{c}(x+(-x)^{a-c+1})-(-1)^d((-x)^{c-d}-(-x)^{a-d})}{(x-1)(-x-1)} \nonumber \\
&+ \frac{(-1)^c(-x-(-x)^{c-b})-(-1)^d((-x)^{b-d+1}-(-x)^{c-d})}{(x-1)(-x-1)}
	\end{align}
Neglecting terms of order $\mathcal{O}(x^{-\min(|a-c|, |b-c|, |a-d|, |b-d|)})$, we are left with
\begin{align}
\sum_{i=a}^b\sum_{j=c}^d (-1)^i x^{-|i-j|}\sgn(j-i) &\approx \frac{(-1)^c x - (-1)^c x}{(x-1)(-x-1)} \nonumber \\
&= 0
\end{align}
We conclude that $\alpha_{12} \approx 0$, as we wished to show. The same argument shows that $\alpha_{21} \approx 0$.

\section{Derivation of Eqs.~(\ref{commid}-\ref{Liid})}
\label{derivklapp}
In this Appendix, we derive Eqs.~(\ref{commid}-\ref{Liid}). We will use two identities in our derivations that are specific to $(k_1, k_2) = (1, 9)$. The first identity is that
\begin{align}
U_{i-1}^{4}U_{i}^{-10}U_{i+1}^{4} =\mathbb{I}. 
\label{U89id}
\end{align}
which follows from  Eq.~(\ref{alg2}). The second identity is that
\begin{align}
4\mathcal{Z}^{-1}_{r-1} - 10\mathcal{Z}^{-1}_{r}+ 4\mathcal{Z}^{-1}_{r+1} = \begin{cases}
1 & r = -1 \\
-1 & r = 1 \\
0 & \text{otherwise}
\end{cases}
\label{z1289id}
\end{align}
which follows from the formula (\ref{eq:zij0}). (Here, we are using the abbreviation $\mathcal{Z}^{-1}_r \equiv \mathcal{Z}^{-1}_{i(i+r)}$). 

We start by proving (\ref{commid}), or more specifically $[K_i, L_j] = [L_i, L_j] = 0$. To prove these relations, consider the quantity $U_i L_j U_i^{-1} L_j^{-1}$. From the commutation algebra (\ref{alg1}) and the identity (\ref{z1289id}), we have
\begin{align}
U_i L_j U_i^{-1} L_j^{-1} &= e^{i 2\pi( 2\mathcal{Z}^{-1}_{2j-i-1} - 5\mathcal{Z}^{-1}_{2j-i}+ 2\mathcal{Z}^{-1}_{2j-i+1})} \mathbb{I} \nonumber \\
&= \begin{cases}
-\mathbb{I} & i = 2j \pm 1 \\
\mathbb{I} & \text{otherwise}
\end{cases}
\label{ULid}
\end{align}
From (\ref{ULid}), it immediately follows that
\begin{align}
[K_i, L_j] = 0
\end{align}
since $K_i$ only contains even powers of $U_i$. Likewise, (\ref{ULid}) implies that
\begin{align}
[L_i, L_j] = 0
\end{align}
since the only odd power in $L_i$, namely $U_{2i}^{-5}$, is not supported on either of the two sites $2j - 1$ or $2j+1$.

We now move on to show (\ref{shiftid}), i.e.~$K_i^2 = K_{i-1}$. To this end, we use the commutation algebra (\ref{alg1}) and the identity (\ref{U89id}) to rewrite $K_{i-1} K_i^{-2}$ as
\begin{align}
	K_{i-1} K_i^{-2}&=e^{-i\varphi_K} \left(U_{i-1}^{4} U_i^{-2}\right) \left(U_{i+1}^{2}U_{i}^{-4}\right) \left(U_{i+1}^{2}U_{i}^{-4}\right) \nonumber \\
	&= e^{i(-\varphi_K + \mu)} U_{i-1}^{4}U_{i}^{-10}U_{i+1}^{4} \nonumber \\
	&=e^{i(-\varphi_K + \mu)} \mathbb{I}
\end{align}
where $\mu = 48 \pi \mathcal{Z}_1^{-1}$.
We conclude that $K_i^2= K_{i-1}$ if we choose
\begin{align}
\varphi_K = \mu = 48 \pi \mathcal{Z}^{-1}_{1}
\label{phikdef}
\end{align}

A corollary of this result is that all the $K_i$'s commute with each other, i.e.
\begin{align}
	[K_i, K_j]=0,
\end{align}
since every $K_i$ can be written as an integer power of every other $K_i$. This establishes the remaining equality in (\ref{commid}).

Next we prove (\ref{Liid}), i.e.~$L_i^2 = \mathbb{I}$. We again use the commutation algebra (\ref{alg1}) and the identity (\ref{U89id}):
\begin{align}
	L_i^2&=e^{i2\varphi_L} \left(U_{2i-1}^{2}U_{2i}^{-5}U_{2i+1}^{2}\right) \left(U_{2i-1}^{2}U_{2i}^{-5}U_{2i+1}^{2}\right) \nonumber \\
	&= e^{i(2\varphi_L + \varsigma)} U_{2i-1}^{4}U_{2i}^{-10}U_{2i+1}^{4} \nonumber \\
	&= e^{i(2\varphi_L + \varsigma)} \mathbb{I}
\end{align}
where
\begin{align}
\varsigma = 2\pi (20\mathcal{Z}^{-1}_{1} - 4 \mathcal{Z}^{-1}_{2})
\end{align}
We conclude that $L_i^2=\mathbb{I}$ if we choose
\begin{align}
\varphi_L = -\frac{\varsigma}{2} = \pi (4 \mathcal{Z}^{-1}_{2} - 20\mathcal{Z}^{-1}_{1})
\label{phildef}
\end{align}
This establishes (\ref{Liid}).


All that remains is to show (\ref{Qid}), i.e.~$K_{2N}^{Q}=\mathbb{I}$. To derive this identity, it is useful to express $K_i$ in terms of $\Gamma_i$. We have
\begin{align}
K_i &= e^{i \varphi_K} U_i^4 U_{i+1}^{-2} \nonumber \\
&= e^{i (\varphi_K + 2 \phi)} e^{i 4 \Gamma_i} e^{- i 2 \Gamma_{i+1}} \nonumber \\
&= e^{i (\varphi_K + 2 \phi - 8 \pi \mathcal{Z}^{-1}_1)} e^{i (4 \Gamma_i -2 \Gamma_{i+1})}
\end{align}
where the last equality follows from the Baker-Campbell-Hausdorff formula. Simplifying the phase in the exponent gives
\begin{align}
\varphi_K + 2 \phi - 8 \pi \mathcal{Z}^{-1}_1 &= 
16 \pi \mathcal{Z}_2^{-1}- 40 \pi \mathcal{Z}_1^{-1} + 4\pi = 0 
\end{align}
where the second equality follows from the identity $2\mathcal{Z}_2^{-1}- 5\mathcal{Z}_1^{-1} = -1/2$, which is a special case of (\ref{z1289id}) above.
Hence 
\begin{align}
	K_i&=e^{i(4\Gamma_i-2\Gamma_{i+1})}
	\label{Kigamma}
\end{align}
without any additional phase factor.

We are now ready to show $K_{2N}^{Q}=\mathbb{I}$, where $Q=\frac{1}{3}(2^{2N}-1)$. First, we note that $Q = \sum_{i=0}^{N-1} 4^i$, so that
\begin{align}
K_{2N}^{Q} &= K_{2N}^{1 + 4 + ... + 4^{N-1}} = \prod_{i=1}^{N} K_{2i} 
\end{align}
Substituting (\ref{Kigamma}) into the right hand side and using the Baker-Campbell-Hausdorff formula gives
\begin{align}
K_{2N}^{Q}=\exp\left(4i\sum_{i=1}^{N}\Gamma_{2i}-2i\sum_{i=1}^{N}\Gamma_{2i+1}+i\varUpsilon\right)
\end{align}
where the phase $\varUpsilon$ is given by
\begin{align}
\varUpsilon=2\pi\sum_{j>i} \left(-4\mathcal{Z}_{2(j-i)-1}^{-1}+10\mathcal{Z}_{2(j-i)}^{-1}-4\mathcal{Z}_{2(j-i)+1}^{-1}\right).
\end{align}
Next we note that each term of the above sum vanishes identically due to the identity (\ref{z1289id}) so $\varUpsilon = 0$. We can then rewrite $K_{2N}^{Q}$ as
\begin{align}
K_{2N}^{Q}&=\exp\left(4i\sum_{i=1}^{N}\Gamma_{2i}-2i\sum_{i=1}^{N}\Gamma_{2i+1}\right) \nonumber \\
 &= \exp \left(i \sum_{i=1}^{2N} \Gamma_i \right) \cdot \exp \left(3i \sum_{i=1}^{2N} (-1)^i \Gamma_i \right) \nonumber \\
\end{align}
Note that there is no Baker-Campbell-Hausdorff phase in the second equality since the two exponents commute with one another, being proportional to $C_{2N+1}$ and $C_{2N+2}$, respectively. Finally, substituting (\ref{globalgamma}) into the right hand side, we obtain the desired result,
\begin{align}
K_{2N}^{Q}&=\mathbb{I}
\end{align}
This proves (\ref{Qid}).

\section{Degeneracy of simultaneous eigenspaces of $K_{2N}, L_1,...,L_N$}
\label{sec:deg}

In this Appendix, we consider the edge theory with $(k_1, k_2) = (1, 9)$ and we compute the degeneracy of each simultaneous eigenspace of $K_{2N}, L_1,...,L_N$. Our main result is that there is a unique eigenstate $|m; \sigma_1,..., \sigma_N\>$ with 
\begin{align*}
K_{2N} |m; \sigma_1,..., \sigma_N\> &= e^{i 2\pi m/Q} |m; \sigma_1,..., \sigma_N\> \nonumber \\
L_i |m; \sigma_1,..., \sigma_N\>&= \sigma_i|m; \sigma_1,..., \sigma_N\>
\end{align*}
for each $m = 0,1,...,Q-1$ and each $\sigma_i = \pm 1$. 

To begin, consider the projector onto this simultaneous eigenpace, which we denote by $P_{m; \sigma_1,...,\sigma_N}$. This projector can be written as a product of $N+1$ spectral projectors -- one for each of the operators $K_{2N}, L_1,..., L_N$:
\begin{align}
P_{m; \sigma_1,...,\sigma_N} = \left(\frac{1}{Q} \sum_{n=0}^{Q-1} e^{-i 2\pi mn/Q} K_{2N}^n  \right) \prod_{i=1}^N \left(\frac{\mathbb{I} + \sigma_i L_i}{2}\right) 
\label{specproj}
\end{align}
Computing the degeneracy of the eigenspace is equivalent to computing the trace of $P_{m; \sigma_1,...,\sigma_N}$. This trace can be obtained by expanding out the above product (\ref{specproj}) into a large sum of terms of the form $K_{2N}^n L_{i_1} \cdots L_{i_k}$. Below we will argue that all of these terms are traceless except for the term consisting of the identity operator $\mathbb{I}$. Once we show this, we will be done since it then follows that
\begin{align}
\text{Tr}(P_{m; \sigma_1,...,\sigma_N}) = \frac{1}{Q} \frac{1}{2^N} \text{Tr}(\mathbb{I}) = 1
\end{align}
Here the second equality follows from the fact that the Hilbert space we are tracing over has a dimension of $D_N = Q \cdot 2^{N}$.

We now explain why every term of the form $K_{2N}^n L_{i_1} \cdots L_{i_k}$ is traceless. We first introduce some notation: for any two operators $O_1, O_2$, we define the bracket
\begin{align}
\lfloor O_1 | O_2 \rfloor \equiv O_1 O_2 O_1^{-1} O_2^{-1}
\end{align}
With this notation, we can now state a useful lemma: 
\begin{lemma}
If $O_1, O_2$ are two operators satisfying
\begin{align}
\lfloor O_1 | O_2 \rfloor = \omega \mathbb{I}, \quad \omega \neq 1
\end{align}
then 
\begin{align}
\text{Tr}(O_1) = \text{Tr}(O_2) = 0. 
\end{align}
\label{tracelemma}
\end{lemma}
To prove this Lemma, note that 
\begin{align*}
O_1 O_2 O_1^{-1} = \omega O_2
\end{align*}
Taking the trace of both sides and using the cyclicity of the trace, we deduce that $\text{Tr}(O_2) = 0$. The same argument shows that $\text{Tr}(O_1) = 0$.

In view of Lemma \ref{tracelemma}, it suffices to find an operator $O$ such that 
\begin{align}
\lfloor O | K_{2N}^n L_{i_1} \cdots L_{i_k} \rfloor = \omega \mathbb{I}
\end{align}
with $\omega \neq 1$. First, we consider the operator $O = U_{2N}^2$. From the commutation algebra (\ref{ULid}) we can see that $U_{2N}^2$ commutes with all the $L_i$'s. Hence
\begin{align}
\lfloor U_{2N}^2 | K_{2N}^n L_{i_1} \cdots L_{i_k} \rfloor &= \lfloor U_{2N}^2 | K_{2N}^n \rfloor \nonumber \\
&= e^{-i 8 n \pi \mathcal{Z}^{-1}_1}
\end{align}
where we are using the abbreviation $\mathcal{Z}^{-1}_r \equiv \mathcal{Z}^{-1}_{i(i+r)}$. Plugging in the formula for $\mathcal{Z}^{-1}_{ij}$ (\ref{eq:zij}) with $(k_1, k_2) = (1, 9)$ and $x=2$, we obtain
\begin{align}
8 n \pi \mathcal{Z}^{-1}_1 = 2 \pi \frac{ n(2^{2N-1} - 2)/3}{Q}
\end{align}
where $Q = \frac{1}{3}(2^{2N} - 1)$. It is easy to check that for any $n=1,...,Q-1$, the above fraction is non-integral and therefore $e^{-i 8 n \pi \mathcal{Z}^{-1}_1} \neq 1$. Hence, by Lemma~\ref{tracelemma} the operator $K_{2N}^n L_{i_1} \cdots L_{i_k}$ is traceless in all of these cases.

The only terms left to consider are those with $n = 0$, i.e. terms of the form $L_{i_1} \cdots L_{i_k}$. Consider any term of this kind that contains $L_i$ but not $L_{i+1}$. In that case, we choose $O = U_{2i+1}$. From the commutation algebra (\ref{ULid}), we can see that $U_{2i+1}$ anticommutes with $L_i$ and $L_{i+1}$ and commutes with all the other $L_i$'s. Therefore, $U_{2i+1}$ anticommutes with the term $L_{i_1} \cdots L_{i_k}$. Applying Lemma~\ref{tracelemma}, we conclude that $L_{i_1} \cdots L_{i_k}$ is traceless. 

At this point, we have shown that all the terms are traceless, except for the term with all the $L_i$'s, i.e. $L_1 L_2 \cdots L_N$. To see that this term is traceless, we note that
\begin{align}
L_1 L_2 \cdots L_N &= L_1^{-1} L_2^{-1} \cdots L_N^{-1} \nonumber \\
&\propto \prod_{i=1}^{2N} U_{2i+1}^{-4} \prod_{i=1}^N U_{2i}^5 \nonumber \\
&\propto (U_1^{-1} U_2 U_3^{-1} \cdots U_{2N})^4 \cdot \prod_{i=1}^N U_{2i} \nonumber \\
&\propto \prod_{i=1}^N U_{2i}
\end{align}
Here, all the proportionality constants are phase factors, and the last line follows from (\ref{alg3}). Now, using (\ref{alg4}), we see that the right hand side is traceless and hence $L_1 L_2 \cdots L_N$ is also traceless. This completes the argument: we have shown that every term in (\ref{specproj}) is traceless except for the term coming from the identity operator $\mathbb{I}$.

\bibliography{references.bib}

\begin{thebibliography}{33}%
\makeatletter
\providecommand \@ifxundefined [1]{%
 \@ifx{#1\undefined}
}%
\providecommand \@ifnum [1]{%
 \ifnum #1\expandafter \@firstoftwo
 \else \expandafter \@secondoftwo
 \fi
}%
\providecommand \@ifx [1]{%
 \ifx #1\expandafter \@firstoftwo
 \else \expandafter \@secondoftwo
 \fi
}%
\providecommand \natexlab [1]{#1}%
\providecommand \enquote  [1]{``#1''}%
\providecommand \bibnamefont  [1]{#1}%
\providecommand \bibfnamefont [1]{#1}%
\providecommand \citenamefont [1]{#1}%
\providecommand \href@noop [0]{\@secondoftwo}%
\providecommand \href [0]{\begingroup \@sanitize@url \@href}%
\providecommand \@href[1]{\@@startlink{#1}\@@href}%
\providecommand \@@href[1]{\endgroup#1\@@endlink}%
\providecommand \@sanitize@url [0]{\catcode `\\12\catcode `\$12\catcode
  `\&12\catcode `\#12\catcode `\^12\catcode `\_12\catcode `\%12\relax}%
\providecommand \@@startlink[1]{}%
\providecommand \@@endlink[0]{}%
\providecommand \url  [0]{\begingroup\@sanitize@url \@url }%
\providecommand \@url [1]{\endgroup\@href {#1}{\urlprefix }}%
\providecommand \urlprefix  [0]{URL }%
\providecommand \Eprint [0]{\href }%
\providecommand \doibase [0]{http://dx.doi.org/}%
\providecommand \selectlanguage [0]{\@gobble}%
\providecommand \bibinfo  [0]{\@secondoftwo}%
\providecommand \bibfield  [0]{\@secondoftwo}%
\providecommand \translation [1]{[#1]}%
\providecommand \BibitemOpen [0]{}%
\providecommand \bibitemStop [0]{}%
\providecommand \bibitemNoStop [0]{.\EOS\space}%
\providecommand \EOS [0]{\spacefactor3000\relax}%
\providecommand \BibitemShut  [1]{\csname bibitem#1\endcsname}%
\let\auto@bib@innerbib\@empty
\bibitem [{\citenamefont {Wen}(1995)}]{wen1995topological}%
  \BibitemOpen
  \bibfield  {author} {\bibinfo {author} {\bibfnamefont {Xiao-Gang}\
  \bibnamefont {Wen}},\ }\bibfield  {title} {\enquote {\bibinfo {title}
  {Topological orders and edge excitations in fractional quantum hall
  states},}\ }\href@noop {} {\bibfield  {journal} {\bibinfo  {journal}
  {Advances in Physics}\ }\textbf {\bibinfo {volume} {44}},\ \bibinfo {pages}
  {405--473} (\bibinfo {year} {1995})}\BibitemShut {NoStop}%
\bibitem [{Note1()}]{Note1}%
  \BibitemOpen
  \bibinfo {note} {This structure consists of a representation of the symmetry
  group $G$ acting on $\protect \mathcal {H}$.}\BibitemShut {Stop}%
\bibitem [{\citenamefont {Kitaev}(2003)}]{KitaevToric}%
  \BibitemOpen
  \bibfield  {author} {\bibinfo {author} {\bibfnamefont {Alexei~Yu}\
  \bibnamefont {Kitaev}},\ }\bibfield  {title} {\enquote {\bibinfo {title}
  {Fault-tolerant quantum computation by anyons},}\ }\href@noop {} {\bibfield
  {journal} {\bibinfo  {journal} {Annals of Physics}\ }\textbf {\bibinfo
  {volume} {303}},\ \bibinfo {pages} {2--30} (\bibinfo {year}
  {2003})}\BibitemShut {NoStop}%
\bibitem [{\citenamefont {Yang}\ \emph {et~al.}(2014)\citenamefont {Yang},
  \citenamefont {Lehman}, \citenamefont {Poilblanc}, \citenamefont
  {Van~Acoleyen}, \citenamefont {Verstraete}, \citenamefont {Cirac},\ and\
  \citenamefont {Schuch}}]{yang2014edgepeps}%
  \BibitemOpen
  \bibfield  {author} {\bibinfo {author} {\bibfnamefont {S.}~\bibnamefont
  {Yang}}, \bibinfo {author} {\bibfnamefont {L.}~\bibnamefont {Lehman}},
  \bibinfo {author} {\bibfnamefont {D.}~\bibnamefont {Poilblanc}}, \bibinfo
  {author} {\bibfnamefont {K.}~\bibnamefont {Van~Acoleyen}}, \bibinfo {author}
  {\bibfnamefont {F.}~\bibnamefont {Verstraete}}, \bibinfo {author}
  {\bibfnamefont {J.~I.}\ \bibnamefont {Cirac}}, \ and\ \bibinfo {author}
  {\bibfnamefont {N.}~\bibnamefont {Schuch}},\ }\bibfield  {title} {\enquote
  {\bibinfo {title} {Edge theories in projected entangled pair state models},}\
  }\href {\doibase 10.1103/PhysRevLett.112.036402} {\bibfield  {journal}
  {\bibinfo  {journal} {Phys. Rev. Lett.}\ }\textbf {\bibinfo {volume} {112}},\
  \bibinfo {pages} {036402} (\bibinfo {year} {2014})}\BibitemShut {NoStop}%
\bibitem [{\citenamefont {Levin}(2018)}]{levin2018talkorder}%
  \BibitemOpen
  \bibfield  {author} {\bibinfo {author} {\bibfnamefont {Michael}\ \bibnamefont
  {Levin}},\ }\href@noop {} {\enquote {\bibinfo {title} {Constraints on order
  and disorder parameters in quantum spin chains and applications},}\ }\bibinfo
  {howpublished}
  {\url{https://www.simonsfoundation.org/event/mps-conference-on-ultra-quantum-matter-ii/}}
  (\bibinfo {year} {2018})\BibitemShut {NoStop}%
\bibitem [{\citenamefont {Ji}\ and\ \citenamefont
  {Wen}(2019)}]{ji2019noninvertible}%
  \BibitemOpen
  \bibfield  {author} {\bibinfo {author} {\bibfnamefont {Wenjie}\ \bibnamefont
  {Ji}}\ and\ \bibinfo {author} {\bibfnamefont {Xiao-Gang}\ \bibnamefont
  {Wen}},\ }\bibfield  {title} {\enquote {\bibinfo {title} {Noninvertible
  anomalies and mapping-class-group transformation of anomalous partition
  functions},}\ }\href {\doibase 10.1103/PhysRevResearch.1.033054} {\bibfield
  {journal} {\bibinfo  {journal} {Phys. Rev. Research}\ }\textbf {\bibinfo
  {volume} {1}},\ \bibinfo {pages} {033054} (\bibinfo {year}
  {2019})}\BibitemShut {NoStop}%
\bibitem [{\citenamefont {Kane}\ and\ \citenamefont
  {Fisher}(1997)}]{kane1997thermal}%
  \BibitemOpen
  \bibfield  {author} {\bibinfo {author} {\bibfnamefont {C.~L.}\ \bibnamefont
  {Kane}}\ and\ \bibinfo {author} {\bibfnamefont {Matthew P.~A.}\ \bibnamefont
  {Fisher}},\ }\bibfield  {title} {\enquote {\bibinfo {title} {Quantized
  thermal transport in the fractional quantum hall effect},}\ }\href {\doibase
  10.1103/PhysRevB.55.15832} {\bibfield  {journal} {\bibinfo  {journal} {Phys.
  Rev. B}\ }\textbf {\bibinfo {volume} {55}},\ \bibinfo {pages} {15832--15837}
  (\bibinfo {year} {1997})}\BibitemShut {NoStop}%
\bibitem [{\citenamefont {Kitaev}\ and\ \citenamefont
  {Kong}(2012)}]{kitaev2012models}%
  \BibitemOpen
  \bibfield  {author} {\bibinfo {author} {\bibfnamefont {Alexei}\ \bibnamefont
  {Kitaev}}\ and\ \bibinfo {author} {\bibfnamefont {Liang}\ \bibnamefont
  {Kong}},\ }\bibfield  {title} {\enquote {\bibinfo {title} {Models for gapped
  boundaries and domain walls},}\ }\href@noop {} {\bibfield  {journal}
  {\bibinfo  {journal} {Communications in Mathematical Physics}\ }\textbf
  {\bibinfo {volume} {313}},\ \bibinfo {pages} {351--373} (\bibinfo {year}
  {2012})}\BibitemShut {NoStop}%
\bibitem [{\citenamefont {Lin}\ and\ \citenamefont
  {Levin}(2014)}]{lin2014stringnet}%
  \BibitemOpen
  \bibfield  {author} {\bibinfo {author} {\bibfnamefont {Chien-Hung}\
  \bibnamefont {Lin}}\ and\ \bibinfo {author} {\bibfnamefont {Michael}\
  \bibnamefont {Levin}},\ }\bibfield  {title} {\enquote {\bibinfo {title}
  {Generalizations and limitations of string-net models},}\ }\href {\doibase
  10.1103/PhysRevB.89.195130} {\bibfield  {journal} {\bibinfo  {journal} {Phys.
  Rev. B}\ }\textbf {\bibinfo {volume} {89}},\ \bibinfo {pages} {195130}
  (\bibinfo {year} {2014})}\BibitemShut {NoStop}%
\bibitem [{\citenamefont {Kong}(2014)}]{kong2014condensation}%
  \BibitemOpen
  \bibfield  {author} {\bibinfo {author} {\bibfnamefont {Liang}\ \bibnamefont
  {Kong}},\ }\bibfield  {title} {\enquote {\bibinfo {title} {Anyon condensation
  and tensor categories},}\ }\href {\doibase
  https://doi.org/10.1016/j.nuclphysb.2014.07.003} {\bibfield  {journal}
  {\bibinfo  {journal} {Nuclear Physics B}\ }\textbf {\bibinfo {volume}
  {886}},\ \bibinfo {pages} {436 -- 482} (\bibinfo {year} {2014})}\BibitemShut
  {NoStop}%
\bibitem [{\citenamefont {Freed}\ and\ \citenamefont
  {Teleman}(2020)}]{freed2020gapped}%
  \BibitemOpen
  \bibfield  {author} {\bibinfo {author} {\bibfnamefont {Daniel~S}\
  \bibnamefont {Freed}}\ and\ \bibinfo {author} {\bibfnamefont {Constantin}\
  \bibnamefont {Teleman}},\ }\bibfield  {title} {\enquote {\bibinfo {title}
  {Gapped boundary theories in three dimensions},}\ }\href@noop {} {\bibfield
  {journal} {\bibinfo  {journal} {arXiv:2006.10200}\ } (\bibinfo {year}
  {2020})}\BibitemShut {NoStop}%
\bibitem [{\citenamefont {Levin}\ and\ \citenamefont
  {Wen}(2005)}]{LevinWenstrnet}%
  \BibitemOpen
  \bibfield  {author} {\bibinfo {author} {\bibfnamefont {Michael~A.}\
  \bibnamefont {Levin}}\ and\ \bibinfo {author} {\bibfnamefont {Xiao-Gang}\
  \bibnamefont {Wen}},\ }\bibfield  {title} {\enquote {\bibinfo {title}
  {String-net condensation: A physical mechanism for topological phases},}\
  }\href {\doibase 10.1103/PhysRevB.71.045110} {\bibfield  {journal} {\bibinfo
  {journal} {Phys. Rev. B}\ }\textbf {\bibinfo {volume} {71}},\ \bibinfo
  {pages} {045110} (\bibinfo {year} {2005})}\BibitemShut {NoStop}%
\bibitem [{Note2()}]{Note2}%
  \BibitemOpen
  \bibinfo {note} {Strictly speaking, this statement is for \protect \emph
  {bosonic} topological phases, but we expect that a similar statement holds in
  the fermionic case.}\BibitemShut {Stop}%
\bibitem [{\citenamefont {Levin}()}]{levinunpublished}%
  \BibitemOpen
  \bibfield  {author} {\bibinfo {author} {\bibfnamefont {Michael}\ \bibnamefont
  {Levin}},\ }\href@noop {} {\bibinfo  {journal} {(unpublished)}\ }\BibitemShut
  {NoStop}%
\bibitem [{\citenamefont {Kapustin}\ and\ \citenamefont
  {Saulina}(2011)}]{kapustin2011topological}%
  \BibitemOpen
\bibfield  {journal} {  }\bibfield  {author} {\bibinfo {author} {\bibfnamefont
  {Anton}\ \bibnamefont {Kapustin}}\ and\ \bibinfo {author} {\bibfnamefont
  {Natalia}\ \bibnamefont {Saulina}},\ }\bibfield  {title} {\enquote {\bibinfo
  {title} {Topological boundary conditions in abelian chern--simons theory},}\
  }\href {\doibase 10.1016/j.nuclphysb.2010.12.017} {\bibfield  {journal}
  {\bibinfo  {journal} {Nuclear Physics B}\ }\textbf {\bibinfo {volume}
  {845}},\ \bibinfo {pages} {393--435} (\bibinfo {year} {2011})}\BibitemShut
  {NoStop}%
\bibitem [{\citenamefont {Levin}(2013)}]{levin2013protected}%
  \BibitemOpen
  \bibfield  {author} {\bibinfo {author} {\bibfnamefont {Michael}\ \bibnamefont
  {Levin}},\ }\bibfield  {title} {\enquote {\bibinfo {title} {Protected edge
  modes without symmetry},}\ }\href {\doibase 10.1103/PhysRevX.3.021009}
  {\bibfield  {journal} {\bibinfo  {journal} {Phys. Rev. X}\ }\textbf {\bibinfo
  {volume} {3}},\ \bibinfo {pages} {021009} (\bibinfo {year}
  {2013})}\BibitemShut {NoStop}%
\bibitem [{\citenamefont {Ganeshan}\ and\ \citenamefont
  {Levin}(2016)}]{ganeshan2016formalism}%
  \BibitemOpen
  \bibfield  {author} {\bibinfo {author} {\bibfnamefont {Sriram}\ \bibnamefont
  {Ganeshan}}\ and\ \bibinfo {author} {\bibfnamefont {Michael}\ \bibnamefont
  {Levin}},\ }\bibfield  {title} {\enquote {\bibinfo {title} {Formalism for the
  solution of quadratic hamiltonians with large cosine terms},}\ }\href
  {\doibase 10.1103/PhysRevB.93.075118} {\bibfield  {journal} {\bibinfo
  {journal} {Phys. Rev. B}\ }\textbf {\bibinfo {volume} {93}},\ \bibinfo
  {pages} {075118} (\bibinfo {year} {2016})}\BibitemShut {NoStop}%
\bibitem [{\citenamefont {Lindner}\ \emph {et~al.}(2012)\citenamefont
  {Lindner}, \citenamefont {Berg}, \citenamefont {Refael},\ and\ \citenamefont
  {Stern}}]{lindner2012fractionalizing}%
  \BibitemOpen
  \bibfield  {author} {\bibinfo {author} {\bibfnamefont {Netanel~H.}\
  \bibnamefont {Lindner}}, \bibinfo {author} {\bibfnamefont {Erez}\
  \bibnamefont {Berg}}, \bibinfo {author} {\bibfnamefont {Gil}\ \bibnamefont
  {Refael}}, \ and\ \bibinfo {author} {\bibfnamefont {Ady}\ \bibnamefont
  {Stern}},\ }\bibfield  {title} {\enquote {\bibinfo {title} {Fractionalizing
  majorana fermions: Non-abelian statistics on the edges of abelian quantum
  hall states},}\ }\href {\doibase 10.1103/PhysRevX.2.041002} {\bibfield
  {journal} {\bibinfo  {journal} {Phys. Rev. X}\ }\textbf {\bibinfo {volume}
  {2}},\ \bibinfo {pages} {041002} (\bibinfo {year} {2012})}\BibitemShut
  {NoStop}%
\bibitem [{\citenamefont {Clarke}\ \emph {et~al.}(2013)\citenamefont {Clarke},
  \citenamefont {Alicea},\ and\ \citenamefont {Shtengel}}]{clarke2013exotic}%
  \BibitemOpen
  \bibfield  {author} {\bibinfo {author} {\bibfnamefont {D.~J.}\ \bibnamefont
  {Clarke}}, \bibinfo {author} {\bibfnamefont {J.}~\bibnamefont {Alicea}}, \
  and\ \bibinfo {author} {\bibfnamefont {K.}~\bibnamefont {Shtengel}},\
  }\bibfield  {title} {\enquote {\bibinfo {title} {Exotic non-abelian anyons
  from conventional fractional quantum hall states},}\ }\href {\doibase
  10.1038/ncomms2340} {\bibfield  {journal} {\bibinfo  {journal} {Nat. Comm.}\
  }\textbf {\bibinfo {volume} {4}},\ \bibinfo {pages} {1348} (\bibinfo {year}
  {2013})}\BibitemShut {NoStop}%
\bibitem [{\citenamefont {Cheng}(2012)}]{cheng2012superconducting}%
  \BibitemOpen
  \bibfield  {author} {\bibinfo {author} {\bibfnamefont {Meng}\ \bibnamefont
  {Cheng}},\ }\bibfield  {title} {\enquote {\bibinfo {title} {Superconducting
  proximity effect on the edge of fractional topological insulators},}\ }\href
  {\doibase 10.1103/PhysRevB.86.195126} {\bibfield  {journal} {\bibinfo
  {journal} {Phys. Rev. B}\ }\textbf {\bibinfo {volume} {86}},\ \bibinfo
  {pages} {195126} (\bibinfo {year} {2012})}\BibitemShut {NoStop}%
\bibitem [{\citenamefont {Kane}\ \emph {et~al.}(1994)\citenamefont {Kane},
  \citenamefont {Fisher},\ and\ \citenamefont
  {Polchinski}}]{kane1994randomness}%
  \BibitemOpen
  \bibfield  {author} {\bibinfo {author} {\bibfnamefont {C.~L.}\ \bibnamefont
  {Kane}}, \bibinfo {author} {\bibfnamefont {Matthew P.~A.}\ \bibnamefont
  {Fisher}}, \ and\ \bibinfo {author} {\bibfnamefont {J.}~\bibnamefont
  {Polchinski}},\ }\bibfield  {title} {\enquote {\bibinfo {title} {Randomness
  at the edge: Theory of quantum hall transport at filling
  \ensuremath{\nu}=2/3},}\ }\href {\doibase 10.1103/PhysRevLett.72.4129}
  {\bibfield  {journal} {\bibinfo  {journal} {Phys. Rev. Lett.}\ }\textbf
  {\bibinfo {volume} {72}},\ \bibinfo {pages} {4129--4132} (\bibinfo {year}
  {1994})}\BibitemShut {NoStop}%
\bibitem [{\citenamefont {Heinrich}\ and\ \citenamefont
  {Levin}(2017)}]{heinrich2017solvable}%
  \BibitemOpen
  \bibfield  {author} {\bibinfo {author} {\bibfnamefont {Chris}\ \bibnamefont
  {Heinrich}}\ and\ \bibinfo {author} {\bibfnamefont {Michael}\ \bibnamefont
  {Levin}},\ }\bibfield  {title} {\enquote {\bibinfo {title} {Solvable models
  for neutral modes in fractional quantum hall edges},}\ }\href {\doibase
  10.1103/PhysRevB.95.205129} {\bibfield  {journal} {\bibinfo  {journal} {Phys.
  Rev. B}\ }\textbf {\bibinfo {volume} {95}},\ \bibinfo {pages} {205129}
  (\bibinfo {year} {2017})}\BibitemShut {NoStop}%
\bibitem [{Note3()}]{Note3}%
  \BibitemOpen
  \bibinfo {note} {See App.~B of Ref.~\protect \rev@citealp
  {levin2013protected}.}\BibitemShut {Stop}%
\bibitem [{\citenamefont {Haldane}(1995)}]{Haldane1995null}%
  \BibitemOpen
  \bibfield  {author} {\bibinfo {author} {\bibfnamefont {F.~D.~M.}\
  \bibnamefont {Haldane}},\ }\bibfield  {title} {\enquote {\bibinfo {title}
  {Stability of chiral luttinger liquids and abelian quantum hall states},}\
  }\href {\doibase 10.1103/PhysRevLett.74.2090} {\bibfield  {journal} {\bibinfo
   {journal} {Phys. Rev. Lett.}\ }\textbf {\bibinfo {volume} {74}},\ \bibinfo
  {pages} {2090--2093} (\bibinfo {year} {1995})}\BibitemShut {NoStop}%
\bibitem [{Note4()}]{Note4}%
  \BibitemOpen
  \bibinfo {note} {More precisely, Eq.~(\ref {heffdiag}) is only guaranteed to
  hold if we make the additional assumption that the matrix $\protect \mathcal
  {Z}_{ij} = \protect \frac {1}{2\pi i} [C_i, C_j]$ has a non-vanishing
  determinant. This property holds for all the systems discussed in this
  paper.}\BibitemShut {Stop}%
\bibitem [{Note5()}]{Note5}%
  \BibitemOpen
  \bibinfo {note} {We will not prove uniqueness in this paper.}\BibitemShut
  {Stop}%
\bibitem [{\citenamefont {Barkeshli}\ \emph {et~al.}(2013)\citenamefont
  {Barkeshli}, \citenamefont {Jian},\ and\ \citenamefont
  {Qi}}]{barkeshli2013twist}%
  \BibitemOpen
  \bibfield  {author} {\bibinfo {author} {\bibfnamefont {Maissam}\ \bibnamefont
  {Barkeshli}}, \bibinfo {author} {\bibfnamefont {Chao-Ming}\ \bibnamefont
  {Jian}}, \ and\ \bibinfo {author} {\bibfnamefont {Xiao-Liang}\ \bibnamefont
  {Qi}},\ }\bibfield  {title} {\enquote {\bibinfo {title} {Twist defects and
  projective non-abelian braiding statistics},}\ }\href {\doibase
  10.1103/PhysRevB.87.045130} {\bibfield  {journal} {\bibinfo  {journal} {Phys.
  Rev. B}\ }\textbf {\bibinfo {volume} {87}},\ \bibinfo {pages} {045130}
  (\bibinfo {year} {2013})}\BibitemShut {NoStop}%
\bibitem [{Note6()}]{Note6}%
  \BibitemOpen
  \bibinfo {note} {Interestingly, $\Delta _{\protect \text {clock}}(m)$
  resembles a discretized version of the Weierstrass function $f(x) = \DOTSB
  \sum@ \slimits@ a^n \protect \qopname \relax o{cos}(b^n \pi x)$ -- a famous
  example of a function that is continuous everywhere but differentiable
  nowhere.}\BibitemShut {Stop}%
\bibitem [{Note7()}]{Note7}%
  \BibitemOpen
  \bibinfo {note} {More generally, one can check that the lowest energy clock
  excitations form a degenerate multiplet of size $4N$, and occur at $m$'s of
  the form $m = \pm 2^k$ for $k=0,1,...,2N-1$.}\BibitemShut {Stop}%
\bibitem [{\citenamefont {Jones}\ and\ \citenamefont
  {Metlitski}(2019)}]{jones20191d}%
  \BibitemOpen
  \bibfield  {author} {\bibinfo {author} {\bibfnamefont {Robert~A}\
  \bibnamefont {Jones}}\ and\ \bibinfo {author} {\bibfnamefont {Max~A}\
  \bibnamefont {Metlitski}},\ }\bibfield  {title} {\enquote {\bibinfo {title}
  {1d lattice models for the boundary of 2d" majorana" fermion spts:
  Kramers-wannier duality as an exact $ z\_2 $ symmetry},}\ }\href@noop {}
  {\bibfield  {journal} {\bibinfo  {journal} {arXiv:1902.05957}\ } (\bibinfo
  {year} {2019})}\BibitemShut {NoStop}%
\bibitem [{\citenamefont {Kane}\ \emph {et~al.}(2002)\citenamefont {Kane},
  \citenamefont {Mukhopadhyay},\ and\ \citenamefont
  {Lubensky}}]{kane2002coupled}%
  \BibitemOpen
  \bibfield  {author} {\bibinfo {author} {\bibfnamefont {C.~L.}\ \bibnamefont
  {Kane}}, \bibinfo {author} {\bibfnamefont {Ranjan}\ \bibnamefont
  {Mukhopadhyay}}, \ and\ \bibinfo {author} {\bibfnamefont {T.~C.}\
  \bibnamefont {Lubensky}},\ }\bibfield  {title} {\enquote {\bibinfo {title}
  {Fractional quantum hall effect in an array of quantum wires},}\ }\href
  {\doibase 10.1103/PhysRevLett.88.036401} {\bibfield  {journal} {\bibinfo
  {journal} {Phys. Rev. Lett.}\ }\textbf {\bibinfo {volume} {88}},\ \bibinfo
  {pages} {036401} (\bibinfo {year} {2002})}\BibitemShut {NoStop}%
\bibitem [{Note8()}]{Note8}%
  \BibitemOpen
  \bibinfo {note} {Alternatively, we could define our electron operators using
  $\Gamma _{2N+2}$ instead of $\Gamma _{2N+1}$; this would give rise to a
  slightly different set of fermion parity-odd operators.}\BibitemShut {Stop}%
\bibitem [{\citenamefont {Newman}(1972)}]{NewmanBook}%
  \BibitemOpen
  \bibfield  {author} {\bibinfo {author} {\bibfnamefont {Morris}\ \bibnamefont
  {Newman}},\ }\href@noop {} {\emph {\bibinfo {title} {Integral Matrices}}}\
  (\bibinfo  {publisher} {Elsevier},\ \bibinfo {year} {1972})\BibitemShut
  {NoStop}%
\end{thebibliography}%
\end{document}